\documentclass[twocolumn,showpacs,preprintnumbers,amsmath,amssymb,aps,nofootinbib]{revtex4}
\usepackage[utf8]{inputenc}
\usepackage{color}
\usepackage{indentfirst}
\usepackage{amsmath,amsthm,amssymb}
\usepackage{graphicx}
\usepackage{setspace}

\begin{document}

\title{Domain wall network as QCD vacuum and the chromomagnetic trap formation under extreme conditions}

\author{ Sergei N.  Nedelko\footnote{nedelko@theor.jinr.ru}, Vladimir E. Voronin\footnote{voronin@theor.jinr.ru}}
\affiliation{ Bogoliubov Laboratory of Theoretical Physics, JINR,
141980 Dubna, Russia }

\begin{abstract}

The ensemble of Euclidean gluon field configurations represented by the domain wall network is considered. A single domain wall is given by the sine-Gordon kink for the angle between chromomagnetic and chromoelectric components of the gauge field.  The  domain wall separates the regions with self-dual and anti-self-dual fields. 
The network  of the domain wall defects  is introduced as a combination of multiplicative and additive superpositions of kinks.  The character of the spectrum and eigenmodes  of  color-charged fluctuations in the presence of the domain wall network is discussed. 
The concept of the confinement-deconfinement transition in terms of the ensemble of  domain wall networks is outlined. 
Conditions for the formation of a stable thick domain wall junction
(the chromomagnetic trap) during heavy ion collisions are discussed, and the spectrum of color charged quasiparticles inside the trap is evaluated. An important observation is the existence of the critical size $L_c$ of the  trap stable against gluon tachyonic modes, which means that deconfinement can occur only in a finite region of space-time in principle. The size $L_c$ is related to the value of gluon condensate $\langle g^2F^2\rangle$.

\end{abstract}


\maketitle

\section{Introduction}

In general, diffusion of the relativized versions of ideas born in  condensed matter and solid state physics to the quantum field theory  has been proven to be extremely fruitful. It was realised long time ago that a complex of problems associated with investigation of the QCD vacuum structure appeared as particularly suitable object in this respect. This paper is focused on the further development of approach to QCD vacuum as a medium describable in terms of statistical ensemble of domain wall networks. This concept plays important role in description of condensed matter systems with rival order and disorder but has been insufficiently explored in application to QCD vacuum.

The identification of the properties of  nonperturbative gauge field configurations relevant to a coherent resolution  of confinement, chiral symmetry breaking, $U_{\rm A}(1)$ and strong CP problems  is an overall task pursued by most approaches to the QCD vacuum structure.

As a  rule, analytical as well as Lattice QCD studies of QCD vacuum structure are focused on localized topological configurations (instantons, monopoles and dyons, vortices) which \textit{via} condensation could be seen as appropriate gauge field configurations responsible for confinement of static color charges and other nonperturbative features of strong interactions. In recent years, three-dimensional configurations akin to domain walls  became popular as well \cite{Ilgenfritz:2007xu, Moran:2008xq, Moran:2007nc,deForcrand:2008aw,deForcrand:2006my,Zhitn}. First of all,  these are the $Z(3)$ domain walls related to the center symmetry 
of the pure Yang-Mills theory~\cite{deForcrand:2008aw} and double-layer domain wall structures in topological charge density~\cite{Zhitn}. Lattice QCD serves as a main source of motivation and verification tool for these studies in pair with the theoretically appealing scenario of static quark confinement in the spirit of the dual Meissner effect  equipped with the Wilson and Polyakov loop criteria. The localized  configurations are characterized by the vanishing ratio of the action to the 4-volume in the infinite volume limit. In this sense, instantons, monopoles, vortices and double-layer domain walls are localized configurations.

A complementary treatment of the above mentioned overall task is based on the investigation of the properties of quantum effective action of QCD.  As in other quantum systems with infinitely many degrees of freedom,  
the global minima of the effective action define the phase structure of QCD.    The identification of global minima in different regimes (high energy density, high baryon density, strong external electromagnetic fields) has highest priority for understanding the phase transformations in hadronic matter. In general, a nontrivial global minimum corresponds to a gauge field with the strength not vanishing at space-time infinity and, hence, extensive action proportional to the  four dimensional space-time volume of the system, unlike the localized configurations. A variety of essentially equivalent statements of the problem in the context of QCD can be found, for instance, in \cite{Minkowski,Pagels,Mink,Leutwyler,NK1,Faddeev0}. Global minima related by  discrete symmetry transformations like CP, Weyl symmetry in the root space of $su(N_{\rm c})$,  center symmetry in particular, is a reason to look for field configurations interpolating between them. First of all, these are domain wall configurations, but also lower dimensional topological defects.
 
There is among others one difference between this treatment  and  approaches based on localized objects:  the last one intends to merge the initially isolated objects (e.g., instanton gas or liquid) while the former collects defects in an initially homogeneous background.  At first sight,  both ways seem to lead to a similar outcome - a class of nonperturbative gluon field configurations with a self-consistent balance of order and disorder which can be characterized, in particular, by nonzero gluon condensate and topological charge density. However, essential   
 disparity can arise since a superposition of localized objects inherits the properties of isolated objects while the superposition of defects in the initially homogeneous ordered background brings some disorder and merely refines the overall properties of the background. For instance, the superposition of infinitely many instantons and anti-instantons is not a configuration with a finite classical action but it maintains the property to have integer-valued topological charge. 
On the contrary, the configuration obtained by implanting infinitely many domain wall defects into the Abelian covariantly constant (anti-)self-dual field can have  any real value of the mean  topological charge density as well as any real value of topological charge fraction per domain~\cite{NK3}.  Both configurations can be seen as lumps of the topological charge density distributed in the Euclidean space-time like in Fig.\ref{Fig:kink_network} or in the lattice
configurations~\cite{Moran:2008xq,Ilgenfritz:2007xu}. However, in the instanton picture each lump carries an integer charge while in the treatment of global minima the charge is any real, irrational, for instance, number. This can have dramatic consequences for the fate of $\theta$ parameter in QCD and the natural resolution of the strong CP-problem~\cite{NK3}.

In the Euclidean formulation, the statement of the problem  starts with the very basic symbol of the functional integral 
\begin{eqnarray*}
 &&Z=N\int\limits_{{\cal F}} DA \exp\{-S[A]\},
 \end{eqnarray*}
 where the functional space ${\cal F}$ is  subject to the condition 
 \begin{eqnarray}
 \label{cond0}
 {\cal F}=\{A: \lim_{V\to \infty} \frac{1}{V}\int\limits_V d^4xg^2F^a_{\mu\nu} (x)F^a_{\mu\nu}(x) =B_{\mathrm vac}^2\}.
\end{eqnarray}
The constant $B_{\mathrm vac}$ is not equal to zero in the general case, which is equivalent to  
nonzero gluon condensate $\langle g^2F^2 \rangle$. 
Condition (\ref{cond0}) singles  out  fields $B_\mu^a$ with the  strength which is constant almost everywhere in $R^4$. It is a necessary requirement to allow gluon condensate to be nonzero. It does not forbid also fields with a finite action since the case $B_{\mathrm vac}=0$ has to be also studied. The dynamics chooses the value of $B_{\mathrm vac}$.
However, the phenomenology of strong interactions has already required nonzero gluon and quark condensates. Hence, they must be allowed in the QCD functional integral from the very beginning. Separation of the long range modes $B_\mu^a$ responsible for gluon condensate and the local fluctuations $Q_\mu^a$ in the background $B_\mu^a$, must be  supplemented by the  gauge fixing condition. The background gauge condition for fluctuations $D(B)Q=0$ is the most natural choice.

Further steps include integration over the fluctuation fields resulting in the effective action for the long-range fields and identification of the minima of this effective action (for more details see \cite{NK1,NK3,NG2011-1}) 
which dominate over the integral in the limit $V\to \infty$ and define the phase structure  of the  system. As soon as minima are identified, this setup defines a principal scheme for self-consistent identification of the class of gauge fields which almost everywhere coincide with the global minima of the quantum effective action.   A treatment of these ``vacuum fields'' in the functional integral
\begin{eqnarray*}
Z &=&N'\int\limits_{{\cal B}}DB \int\limits_{{\cal Q}} DQ \det[D(B)D(B+Q)]
 \\
 &&\times\delta[D(B)Q] \exp\{-S[B+Q]+S[B]\}.
\end{eqnarray*}
 must be nonperturbative. The fields $B_\mu^a\in{\cal B}$ are subject to condition (\ref{cond0}) with the fixed vacuum value of the condensate $B_{\mathrm vac}^2$. The condensate plays the role of the scale parameter of QCD to be identified from the hadron phenomenology. 
The fluctuations $Q$ in the background of the vacuum fields can be seen as perturbations. 

The homogeneous fields with the domain wall defects  are the most natural and simplest  example of  gluon configurations which are homogeneous almost everywhere in $R^4$ and satisfy the basic condition Eq.(\ref{cond0}). Basic argumentation in favour of the Abelian (anti-)self-dual homogeneous fields as global minima of the effective action originates from  papers
\cite{ Minkowski, Leutwyler,Pagels,Woloshin,Pawlowski,NG2011}.

Within the Ginzburg-Landau approach to the effective action the domain wall is described simply by the sine-Gordon kink for the angle between chromomagnetic and chromoelectric components of the gluon field~\cite{NG2011}. This  kink configuration can be seen as either  Bloch or N\'eel domain wall separating the regions with  self-dual and anti-self-dual Abelian gauge fields. On the domain wall the gluon field is Abelian with orthogonal to each other chromomagnetic and chromoelectric fields.  We shall not repeat here arguments leading to this conclusion  but just refer to papers~\cite{NG2011,NG2011-1} where a more detailed discussion can be found.  Group theoretical analysis of the Weyl symmetry and subgroup embeddings behind the domain wall formation in the effective gauge theories is given in a recent paper~\cite{George:2012sb}. 

It should be also mentioned that the model of confinement, chiral symmetry breaking and hadronization based on the  dominance of the gluon fields which are (anti-)self-dual Abelian almost everywhere demonstrated high phenomenological performance~\cite{EN,NK1,NK3,NK4}.  

The purpose of the present paper is to evolve the approach outlined in article~\cite{NG2011} in two respects: explicit analytical construction of the domain wall network in $R^4$  through a combination of additive and multiplicative superpositions of kinks,  and refining the spectrum and eigenmodes of  the color charged scalar, spinor and vector fields in the background of a domain wall. 
In particular the spectrum of quasiparticles inside the thick domain wall junction is evaluated.

It is shown that the standard methods of the sine-Gordon model~\cite{Vachaspati} allow one to generate various domain wall networks. 
The eigenvalues and eigenfunctions are found for the Laplace operator in the background  of a single infinitely thin domain wall. In this case, the eigenvalue problem has to be solved separately in the bulk of $R^4$
and on the 3-dimensional hyperplane of the wall. The continuity of the charge current through the wall is required together with the square integrability of the bulk eigenfunctions. For the infinitely thin wall the bulk eigenfunction possesses the purely discrete spectrum which coincides with the spectrum for the case of a homogeneous (anti-)self-dual field without a kink defect. The eigenfunctions differ in a certain way but have the same harmonic oscillator type as the ones in the absence of the kink defect. These modes describe confined color charged fields.  The eigenfunctions localized on the wall have  continuous spectrum with the dispersion relation of  charged quasi-particles. This confirms qualitative conjectures  of \cite{NG2011,NG2011-1}. 

It is argued that thick domain wall junction may be formed
during heavy ion collisions and play the role of a trap for charged
quasi-particles. Confinement is lost inside the trap of a finite size. There exists a critical size of the stable trap, beyond which the emerging tachyonic gluon modes destroy it. 

The paper is organized as follows. Section~II is devoted to the domain wall network construction. The spectrum of scalar color charged field in the background of infinitely thin domain wall is discussed in section~III. In the fourth section we discuss  the chromomagnetic trap formation and evaluate the spectrum and eigenmodes of the color charged scalar, vector and spinor quasiparticles inside the trap.

\section{Nonzero gluon condensate $\langle g^2F^2\rangle$ and domain wall network in QCD vacuum}

\begin{figure}[h!]
 \centerline{\includegraphics[width=75mm,angle=0]{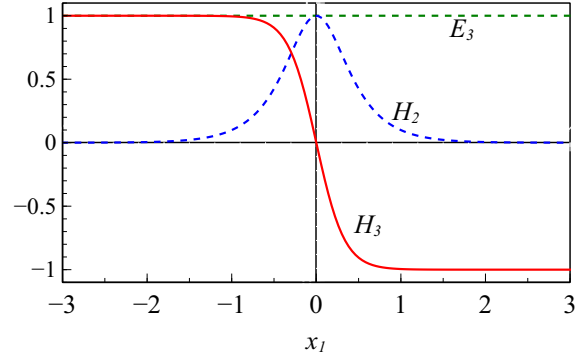}} 
%\vspace{-5mm}
\centerline{\includegraphics[width=70mm,angle=0]{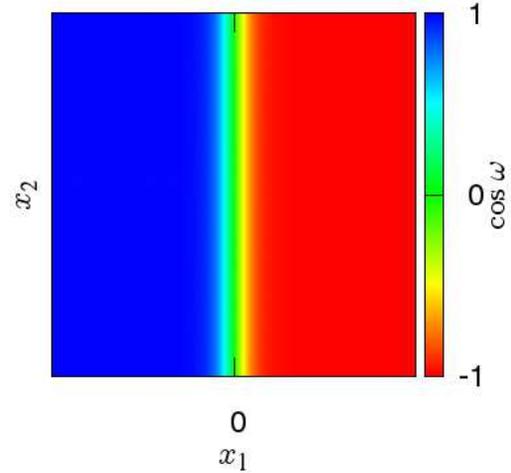}}
    \caption{Kink profile in terms of the components of the chromomagnetic and chromoelectric field strengths (upper plot), and a two-dimensional slice for the topological charge density  in the presence of a single kink measured in units of $g^2F^b_{\alpha\beta}F^b_{\alpha\beta}$ (lower plot). 
Here $\omega$ is the angle between the chromomagnetic and chromoelectric fields, $\cos\omega=F^a_{\mu\nu}\tilde F^a_{\mu\nu}/F^b_{\alpha\beta}F^b_{\alpha\beta}$.  The three-dimensional planar domain wall separates the four-dimensional regions filled with the self-dual (blue color) and anti-self-dual (red color) Abelian covariantly constant gluon fields. The chromomagnetic and chromoelectric fields are orthogonal to each other inside the wall (green color). }
 \label{Fig:single_kink}
\end{figure}

The calculation of the effective quantum action for the Abelian (anti-)self-dual homogeneous gluon field within the functional renormalization group  approach~\cite{Pawlowski} has indicated that this configuration is a serious candidate for the role of global minimum of QCD effective action and has enhanced the older one-loop results \cite{Minkowski,Leutwyler,Pagels}.  The functional  RG result also supported conclusions of~\cite{NK1,NG2011} based on the Ginzburg-Landau type effective Lagrangian of the form
\begin{eqnarray}
    \mathcal{L}_{\mathrm{eff}} &=& - \frac{1}{4\Lambda^2}\left(D^{ab}_\nu F^b_{\rho\mu} D^{ac}_\nu F^c_{\rho\mu} + D^{ab}_\mu F^b_{\mu\nu} D^{ac}_\rho F^c_{\rho\nu }\right)  
    \nonumber\\
    &-&U_{\mathrm{eff}}   
    \nonumber \\
U_{\mathrm{eff}}&=&\frac{\Lambda^4}{12} {\rm Tr}\left(C_1\breve{ f}^2 + \frac{4}{3}C_2\breve{ f}^4 - \frac{16}{9}C_3\breve{ f}^6\right),
\label{ueff}
\end{eqnarray}
where $\Lambda$ is a scale of QCD related to gluon condensate, $\breve f=\breve F/\Lambda^2$, and  
\begin{eqnarray*}
 && D^{ab}_\mu = \delta^{ab} \partial_\mu - i\breve{ A}^{ab}_\mu = \partial_\mu - iA^c_\mu {(T^c)^{ab}},
\\
 && F^a_{\mu\nu} = \partial_\mu A^a_\nu - \partial_\nu A^a_\mu - if^{abc} A^b_\mu A^c_\nu,
\\
 && \breve{ F}_{\mu\nu} = F^a_{\mu\nu} T^a,\ \ \ T^a_{bc} = -if^{abc}
\\
 && {\rm Tr}\left(\breve{ F}^2\right) = \breve{ F}^{ab}_{\mu\nu}\breve{ F}^{ba}_{\nu\mu} = -3 F^a_{\mu\nu}F^a_{\mu\nu} \leq 0,
\\ && C_1>0, \ C_2>0, \ C_3 > 0.
\end{eqnarray*}
Detailed discussion of this expression can be found in \cite{NG2011}. Here it should be noted that all symmetries of 
QCD are respected and the signs of the constants are chosen so that the action is bounded from below and its minimum corresponds to the fields with nonzero strength, i.e. $F^2\not=0$ at the minimum. Thus, an important input is the existence of the nonzero gluon condensate. By inspection, one gets as an output twelve (for $SU(3)$) global degenerate discrete minima. The minima are achieved for covariantly constant Abelian (anti-)self-dual fields
\begin{eqnarray*}
\breve{ A}_{\mu}  = -\frac{1}{2}\breve{ n}_k F_{\mu\nu}x_\nu, \, \tilde F_{\mu\nu}=\pm F_{\mu\nu}
\end{eqnarray*}
where the matrix $\breve{n}_k$ belongs to the Cartan subalgebra of $su(3)$
\begin{eqnarray}
 \breve n_k &=& T^3\ \cos\left(\xi_k\right) + T^8\ \sin\left(\xi_k\right),
\nonumber\\
\xi_k&=&\frac{2k+1}{6}\pi, \, k=0,1,\dots,5.
\label{HLxik}
\end{eqnarray}
The values $\xi_k$ correspond to the boundaries of the Weyl chambers in the root space of $su(3)$. The minima are connected by the discrete parity and Weyl transformations, which indicates that the system is prone to existence of solitons (in real space-time) and kink configurations (in Euclidean space). Below we shall concentrate on the simplest 
configuration -- kink interpolating between self-dual and anti-self-dual Abelian vacua. If the angle $\omega$ between chromoelectric and chromomagnetic fields is allowed to deviate from the constant vacuum value and all other parameters  are fixed to the vacuum values, then the Lagrangian takes the form 
\begin{eqnarray*} 
\label{SG}
\mathcal{L}_{\textrm{eff}} &=& -\frac{1 }{2}\Lambda^2 b_{\textrm{vac}}^2 \partial_\mu \omega \partial_\mu \omega 
\\
&-& b_{\textrm{vac}}^4  \Lambda^4 \left(C_2+3C_3b_{\textrm{vac}}^2 \right){\sin^2\omega},
\end{eqnarray*}
with the corresponding sine-Gordon equation
\begin{eqnarray*}
    \partial^2\omega = m_\omega^2 \sin 2\omega,\ \   m_\omega^2 = b_{\textrm{vac}}^2  \Lambda^2\left(C_2+3C_3b_{\textrm{vac}}^2 \right),
\end{eqnarray*}
and the standard kink solution
\begin{equation*}
 \omega(x_\mu) = 2\ {\rm arctg} \left(\exp(\mu x_\mu)\right)
\label{sakink}
\end{equation*}
interpolating between  $0$ and $\pi$. Here $x_\mu$ stays for one of the four Euclidean coordinates. The kink describes a planar domain wall between the regions with almost homogeneous Abelian self-dual and anti-self-dual gluon fields. Chromomagnetic and chromoelectric fields are orthogonal to each other on the wall, see Fig.\ref{Fig:single_kink}.  Far from the wall, the topological charge density is constant, its absolute value is equal to the value of the gluon condensate.  The topological charge density vanishes on the wall.
 The upper plot shows the profiles of the components of the chromomagnetic and chromoelectric fields corresponding to the Bloch domain wall -- the chromomagnetic field flips in the direction parallel to the wall plane. 

\begin{figure}[h!]
\begin{center}
\includegraphics[width=70mm]{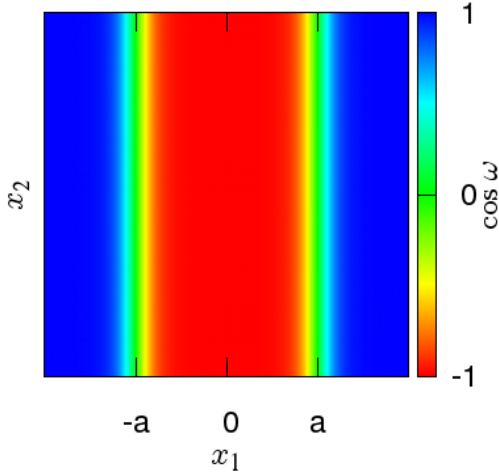}
\parbox[t]{0.47\textwidth}{\caption{Two-dimensional slice of a multiplicative superposition of two kinks.}\label{multsupkink}}
\end{center}
\end{figure}

\begin{figure}[h!]
\begin{center}
\includegraphics[width=75mm]{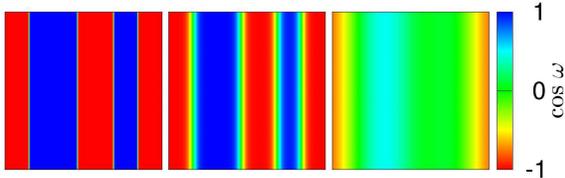}
\parbox[t]{0.47\textwidth}{\caption{Two-dimensional slice of the layered topological charge distribution in $R^4$ according to Eq.(\ref{layered}). The action density is equal to the same nonzero constant value for all three configurations. The LHS plot represents a configuration with infinitely thin planar Bloch domain wall defects, which is the Abelian homogeneous (anti-)self-dual field almost everywhere in $R^4$, characterized by the nonzero absolute value of the topological charge density almost everywhere proportional to the value of the action density. 
The most RHS plot shows the opposite case of very thick kink network. Green color corresponds to the gauge field with infinitesimally small topological charge density.  Most LHS configuration is confining (only colorless hadrons can be excited) while most RHS one supports the color charged quasiparticles as elementary excitations.}
\label{Fig:layered}}
\end{center}
\end{figure}

\begin{figure}[h!]
 \centerline{\includegraphics[width=70mm,angle=0]{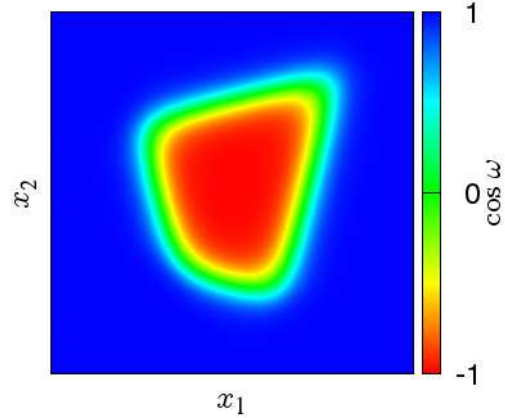}}
    \caption{A two-dimensional slice of the four-dimensional lump of anti-self-dual field in the background of the self-dual configuration. The domain wall surrounding the lump in the four-dimensional space is given by the multiplicative superposition of eight kinks as it is defined by Eq.(\ref{asd_lump}).  }
 \label{Fig:asd_lump}
\end{figure}

The domain wall network can be now constructed by the standard methods~\cite{Vachaspati}. 
Let us denote the general kink configuration as
\begin{equation*}
 \zeta(\mu_i,\eta_\nu^{i}x_\nu-q^{i})=\frac{2}{\pi}\arctan\exp(\mu_i(\eta_\nu^{i}x_\nu-q^{i})),
\end{equation*}
where $\mu_i$ is the inverse width of the kink, $\eta_\nu^{i}$ is a normal vector to the plane of the wall,
$q^{i}=\eta_\nu^{i}x^{i}_\nu$ with $x^{i}_\nu$ - coordinates of the wall.
The topological charge density for the multiplicative superposition of two kinks with the normal vectors  
anti-parallel to each other
\begin{equation*}
\omega(x_1)=\pi\zeta(\mu_1,x_1-a_1)\zeta(\mu_2,-x_1-a_2)
\end{equation*}
 is shown in Fig.\ref{multsupkink}. The additive superposition of infinitely many pairs
\begin{equation}
\label{layered}
\omega(x_1)=\pi\sum\limits_{j=1}^{\infty}\zeta(\mu_j,x_1-a_j)\zeta(\mu_{j+1},-x_1-a_{j+1})
\end{equation}
gives a layered topological charge structure in $R^4$, Fig.\ref{Fig:layered}. 

\begin{figure}[h!]
{\includegraphics[width=25mm,angle=0]{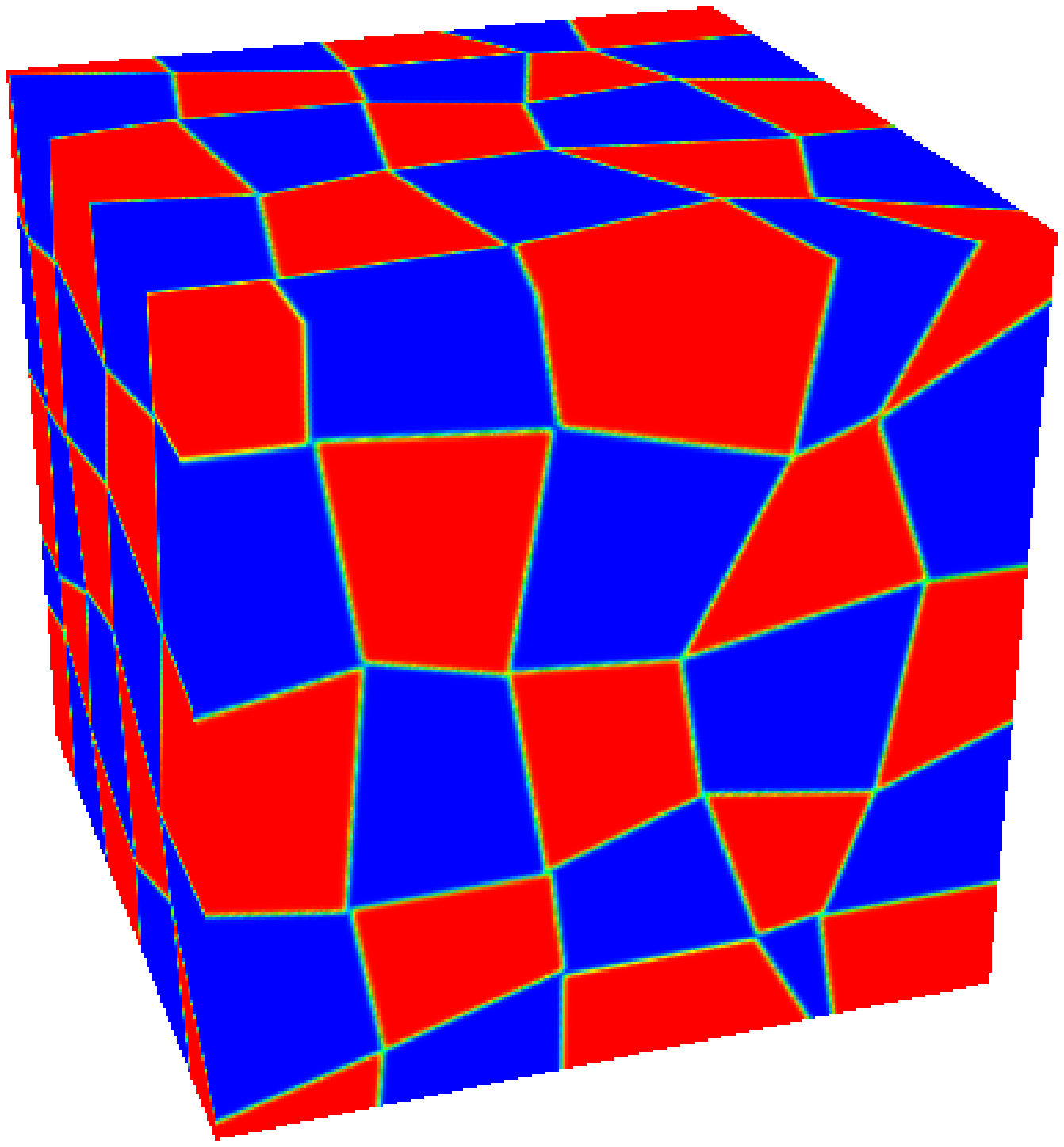}}
 {\includegraphics[width=25mm,angle=0]{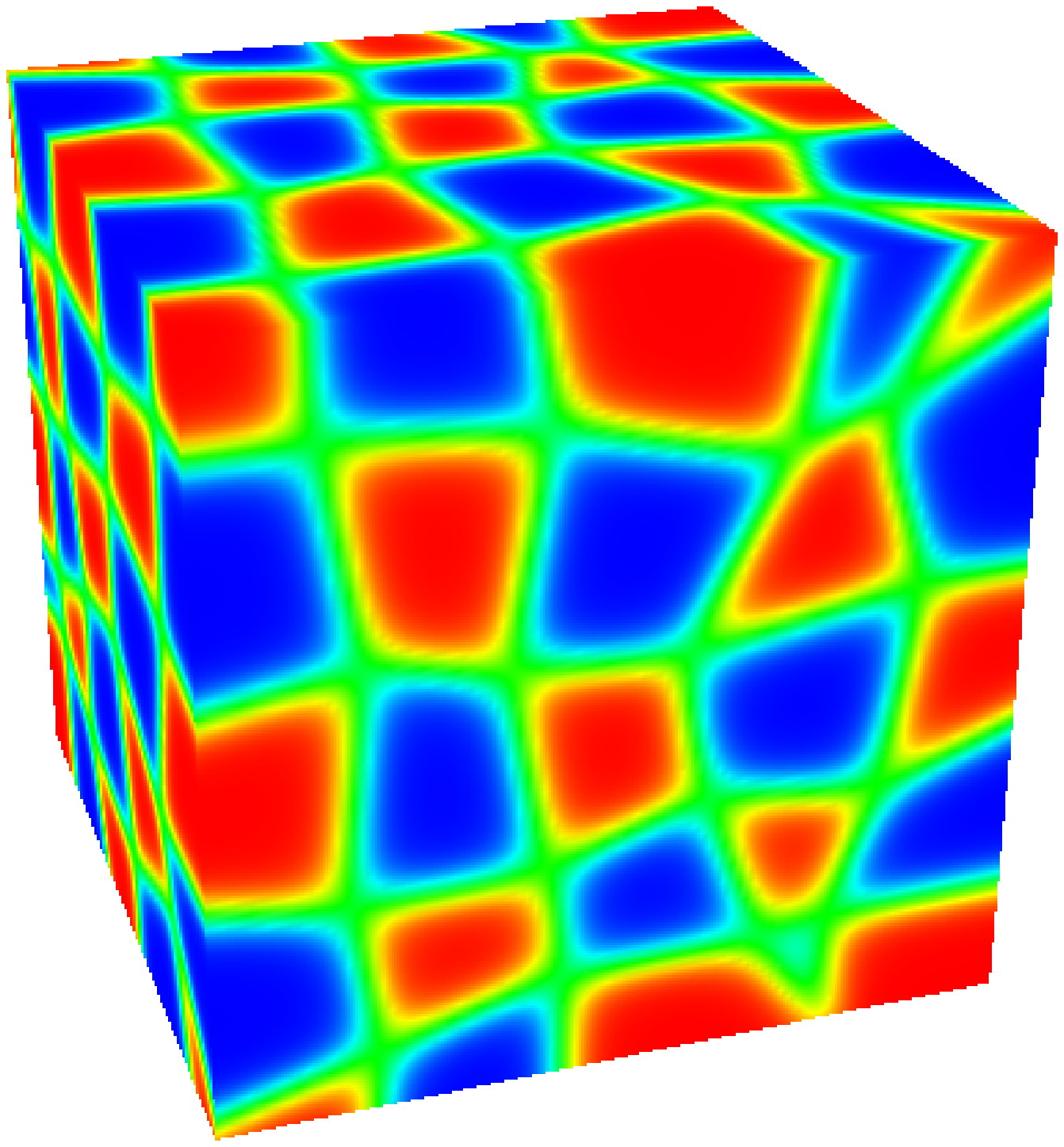}}
{\includegraphics[width=25mm,angle=0]{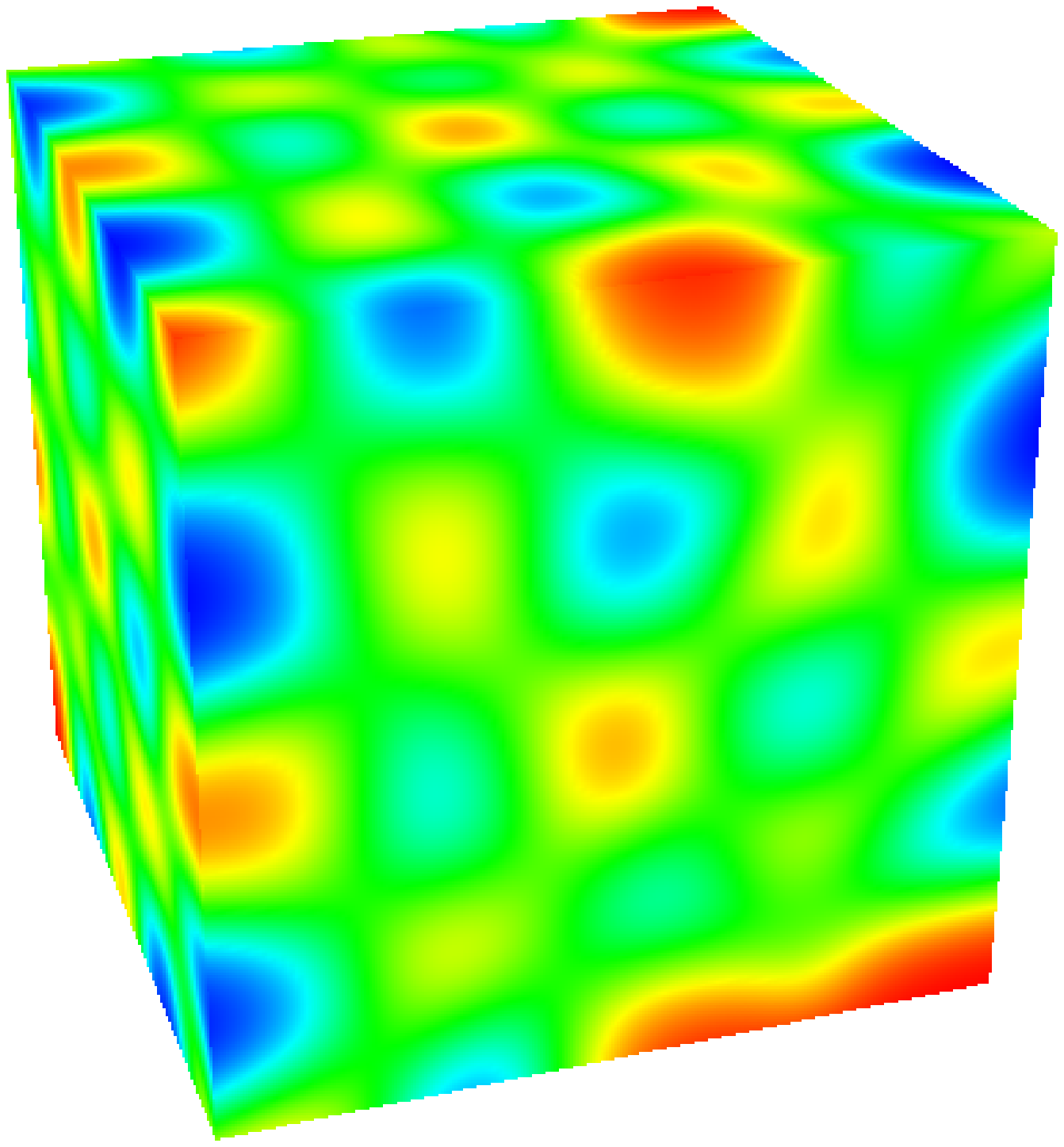}}
    \caption{Three-dimensional slices of the kink network - additive superposition of numerous four-dimensional lumps as it is given by 
    Eq.~\eqref{kink_network}. The correspondence of colors to the character of the configuration is the same as in Fig.~\ref{Fig:layered}.  }
 \label{Fig:kink_network}
\end{figure}

Formally, one may try to go further and consider the product
\begin{equation}
 \label{asd_lump}
\omega(x)=\pi\prod_{i=1}^k \zeta(\mu_i,\eta_\nu^{i}x_\nu-q^{i}).
\end{equation}
For an appropriate choice of normal vectors $\eta^i$ this superposition represents a lump of anti-self-dual field in the background of the self-dual one, in two, three and four dimensions for $k=4,6,8$, respectively. The case $k=8$ is illustrated in Fig.\ref{Fig:asd_lump}. 
The general kink network is then given by the additive superposition of lumps (\ref{asd_lump})
\begin{equation}
 \label{kink_network}
\omega=\pi\sum_{j=1}^{\infty}\prod_{i=1}^k \zeta(\mu_{ij},\eta_\nu^{ij}x_\nu-q^{ij}).
\end{equation}
The correponding topological charge density is shown in 
Fig.~\ref{Fig:kink_network}.  
This figure as well as the LHS of Fig.~\ref{Fig:layered}  represents the configuration with infinitely thin domain wall defects, that is the Abelian homogeneous (anti-)self-dual field almost everywhere in $R^4$ characterized by the nonzero absolute value of the topological charge density which is constant  and proportional to the value of the action density almost everywhere. 

The most RHS plots in Figs.~\ref{Fig:layered} and  \ref{Fig:kink_network} show the opposite case of the network composed of very thick kinks. Green color corresponds to the gauge field with an infinitesimally small topological charge density. Study of the spectrum of colorless and  color charged fluctuations indicates that the LHS configuration is expected to be confining (only colorless hadrons can be excited as particles) while the RHS  one (crossed orthogonal field) supports the color charged quasiparticles as the elementary excitations. It is expected that the RHS configuration can be triggered by external electromagnetic fields ~\cite{NG2011-1,D'Elia:2012zw,Bali:2013esa}. Strong electromagnetic fields   emerge in relativistic heavy ion collisions~\cite{Skokov:2009qp,toneev,Warringa}.  Even after switching off the external electromagnetic field the nearly pure chromomagnetic vacuum configuration (RHS Fig.\ref{Fig:kink_network}) can support strong anisotropies \cite{Tuchin:2013ie} and, in particular, influence the chiral symmetry realization in the collision region \cite{Fukushima:2012kc}.
More detailed consideration of the spectrum of elementary color charged excitations at the domain wall junctions (the green regions) is given in the section~\ref{trap}.  

A comment on representation of the domain wall network in terms of the vector potential is in order. The domain wall network constructed in this section relies on the separation of the  Abelian part from the general gauge field. The vector potential representation can be easily realized for the planar Bloch domain wall and their layered superposition, Fig.~\ref{Fig:layered}. The same is true also for the interior of a thick domain wall junction, where field is almost homogeneous. The description of the domain walls in the general network Fig.~\ref{Fig:kink_network}  in terms of the vector potential requires application of the gauge field parametrization  suggested in a series of papers by Y.M.~Cho \cite{Cho1,Cho2},  S. Shabanov~\cite{shabanov1, shabanov2},   L.D.~Faddeev and  A. J. Niemi \cite{Faddeev} and, recently, by K.-I. Kondo \cite{Kondo}. In this parameterization the Abelian part ${\hat V}_\mu (x)$ of the gauge field ${\hat A}_\mu (x)$ is
 separated manifestly,
\begin{eqnarray*}
\label{eqnsFaddeevNiemKondo} 
 {\hat A}_\mu (x) &=& {\hat V}_\mu (x) + {\hat X}_\mu (x), \, 
{\hat V}_\mu (x) = {\hat B}_\mu (x) + {\hat C}_\mu (x), \\
 {\hat B}_\mu (x) &=& [n^aA^a_\mu (x)]\hat{n} (x)=B_\mu(x)\hat{n}(x), \nonumber \\
 {\hat C}_\mu (x) &=& g^{-1}\partial_\mu \hat{n}(x)\times \hat{n}(x), \nonumber\\
 {\hat X}_\mu (x) &=& g^{-1}{\hat n}(x) \times \left( \partial_\mu {\hat n}(x) + g {\hat A}_\mu (x) \times {\hat n}(x) \right), \nonumber
\end{eqnarray*}
where ${\hat A}_\mu (x) = A^a_\mu (x) t^a$, ${\hat n} (x) = n_a (x) t^a$,   $n^a n^a = 1$, and
\begin{eqnarray*}
{\partial_\mu\hat n}\times {\hat n} = i f^{abc}\partial_\mu n^a n^b t^c,\,
      \, [t^a,t^b]=if^{abc}t^c.
\end{eqnarray*}
The field ${\hat V}_\mu $ is seen as the Abelian field in the sense that $[{\hat V}_\mu (x),{\hat V}_\nu (x)]=0$.
The color vector field  $n^a(x)$ may be used for detailed description of the thin domain wall junctions in general case. This issue is beyond the scope of the present paper and will be considered elsewhere. 

\section{Charged field fluctuations in the background of a planar domain wall}

\subsection{Boundary condition}

In this section we study the spectrum of color charged field fluctuations in the background of a single planar domain wall of the Bloch type.

The best thing to do would be to solve the eigenvalue problem for the kink of the finite width. 
However, the problem turns out to be not that simple. Let us consider the problem for the scalar field in the adjoint representation, that is just the Faddeev-Popov ghost field in the background gauge.  The quadratic part of the action for the scalar field in the background field of a planar kink with the finite width  placed at $x_1=0$ looks like
\begin{eqnarray}
\label{action_sc}
S[\Phi]&=&-\int d^4x (D_\mu\Phi)^\dagger(x)D_\mu\Phi(x)
\\ \nonumber
&=&\int d^4x \Phi^\dagger(x)D^2\Phi(x),
\\
  \nonumber
D_\mu&=&\partial_\mu+i\breve B_\mu, \, \breve B_\mu=-\breve n B_\mu(x).
\end{eqnarray}
Here $\breve n$ is the constant color matrix, $B_\mu$ is the vector potential for the planar Bloch domain wall.
For our purposes the most convenient gauge for $B_\mu$ is
\begin{eqnarray}
\label{gauge1}
&& B_1=H_2(x_1)x_3+H_3(x_1)x_2,
\\ \nonumber
&& B_2=B_3=0,\quad B_4=-Bx_3,
\\ \nonumber
&&H_2=B\sin\omega(x_1), \,H_3=-B\cos\omega(x_1), 
\\ \nonumber
&&\omega(x_1)=2\ {\rm arctg} \exp\mu x_1.
\end{eqnarray}

A kink with the finite width is a regular everywhere in $R^4$ function, the scalar field is assumed to be a continuous  square integrable function. Integration by parts in Eq.(\ref{action_sc}) does not generate surface terms either at infinity or at the location of the kink. However, there is a peculiarity related to the chosen gauge of the background field. According to Eq.(\ref{gauge1}),
\begin{eqnarray}
 D^2&=& \tilde D^2+i\partial_\mu\breve B_\mu, 
\label{covder1}\\
\tilde D^2&=& \partial^2+2i\breve B_\mu\partial_\mu -i\breve B_\mu \breve B_\mu
\nonumber\\
&=& (\partial_1-i\breve n H_2(x_1)x_3-i\breve nH_3(x_1)x_2)^2 
\nonumber\\
&& + \partial_2^2 +\partial_3^2 + (\partial_4+i\breve n Bx_3)^2 -i\partial_1B_1
\nonumber\\
\partial_\mu\breve B_\mu&=&-\breve n H^\prime_2(x_1)x_3-\breve n H^\prime_3(x_1)x_2.
\label{sing_gauge}
\end{eqnarray}
The action can be written as
\begin{eqnarray}
S[\Phi]&=&\int d^4x \Phi^\dagger(x)\tilde D^2\Phi(x)
\label{surf_gauge}\\
&-&i\int d^4x \Phi^\dagger(x)\breve n\Phi(x) \left[H^\prime_2(x_1)x_3+H^\prime_3(x_1)x_2\right] .  
\nonumber
\end{eqnarray}
It should be noted that the integral in the second line is equal to zero if 
$\Phi^\dagger(x)\breve n\Phi(x)$ is an even function of $x_2$ and $x_3$. 

The structure of $D^2$ in Eq.(\ref{covder1}) is quite complicated.  In the eigenvalue problem the variables can hardly be separated in the case of the finite width of the kink. The problem becomes much simpler and tractable in the limit of the infinitely thin domain wall $\mu\to\infty$. This limit brings discontinuity into the background field and thus creates a sharp boundary -- the hyperplane of the domain wall. In such a situation one has to solve the problem in the bulk and on the wall and match the solutions according to some appropriate conditions. For our choice of the kink location there are three regions to be studied:  $x_1<0$ with the self-dual field $B_\mu$, $x_1>0$ with the anti-self-dual field, and $x_1=0$ with the chromomagnetic and chromoelectric fields orthogonal to each other. Conditions imposed onto the eigenmodes of color charged fields on the sharp wall can be obtained from the requirement of preservation of the properties of eigenmodes for finite $\mu$ as far as they can be identified. The continuity of the normal to the wall component of the total (through the whole hypersurface of the wall) charged current offers a reliable guiding principle for identification of the matching conditions. Continuity of the total current means that the surface terms do not appear under integration by parts in the action,
\begin{eqnarray}
\label{current_cont_sc}
&&\lim_{\varepsilon\to 0}\left[J_1(\varepsilon)-J_1(-\varepsilon)\right]=0,
\\
\nonumber
&&J_\mu(x_1)=\int d^3x \Phi^\dagger(x)D_\mu\Phi(x),
\\
\nonumber
&&d^3x=dx_2dx_3dx_4.
\end{eqnarray}
Moreover, this requirement restricts the form of the eigenfunctions in such a way that the surface terms associated with the gauge dependent delta-function singularuties in $\partial_\mu \breve B_\mu$, Eq.(\ref{sing_gauge}) vanish as well.  

\begin{figure}[h!]
 \centerline{\includegraphics[width=75mm,angle=0]{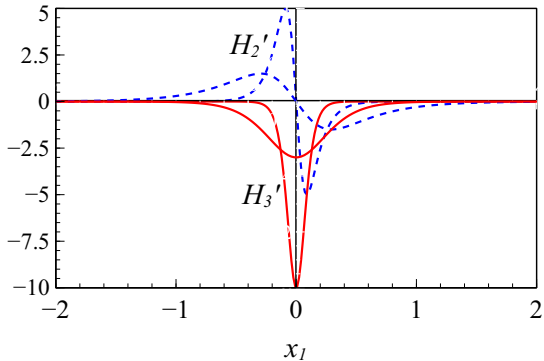}} 
%\vspace{-20mm}
%\centerline{\includegraphics[width=70mm,angle=0]{single_kink}}
    \caption{Derivatives of the components of the chromomagnetic field are plotted for two values of the width parameter $\mu/\sqrt{B}=3,10$.The coordinate $x_1$ is given in units of $1/\sqrt{B}$. In the limit of the infinitely thin domain wall ($\mu/\sqrt{B}\to\infty$) the derivatives develop the delta-function singularities at the location of the wall. }
 \label{Fig:der_single_kink}
\end{figure}

\subsection{Confined fluctuations in the bulk}

Let us consider the  eigenvalue problem
\begin{eqnarray*}
\label{scalar1}
&&-\tilde D^2\Phi=\lambda\Phi.
\end{eqnarray*}
for the functions square integrable in $R^4$ and satisfying the condition (\ref{current_cont_sc}).
For all $x_1\not=0$ the operator $\tilde D^2$ takes the form
\begin{eqnarray}
\label{covder2}
\tilde D^2&=& (\partial_1 \pm i\breve n B x_2)^2 
\nonumber\\
&& + \partial_2^2 +\partial_3^2 + (\partial_4+i\breve n Bx_3)^2 
\nonumber
\end{eqnarray}
where plus corresponds to the anti-self-dual configuration and minus is for the self-dual one.  
By inspection one can see that the eigenfunctions satisfy the relation
\begin{eqnarray}
 \label{conditions}
\Phi^{(+)}(x_1,x_\perp)=\Phi^{(-)}(-x_1,x_\perp),
\end{eqnarray}
where $(\pm)$ denotes the duality of the background field for a given $x_1$. 

Respectively, the square integrable solutions are
\begin{eqnarray*}
\label{eigenf_sc}
\Phi^{(\pm)}_{kl}(x)&=&\phi^{(\pm)}_k(x_1,x_2)\chi_l(x_3,x_4)
\\
\phi^{(\pm)}_k(x_1,x_2)&=&\int dp_1 f(p_1)e^{\pm ip_1x_1-\frac{1}{2}|\breve n|B(x_2+p_1/|\breve n|B)^2}
\nonumber\\
&\times& H_k\left(\sqrt{|\breve n|B}\left[x_2+\frac{p_1}{|\breve n|B}\right]\right)
\nonumber\\
\chi_k(x_3,x_4)&=&\int dp_4 g(p_4)e^{ip_4x_4-\frac{1}{2}|\breve n|B(x_3+p_4/|\breve n|B)^2 }
\nonumber\\
&\times& H_l\left(\sqrt{|\breve n|B}\left[x_3+\frac{p_4}{|\breve n|B}\right]\right),
\nonumber
\end{eqnarray*}
where $H_m$ are the Hermite polynomials. The eigenvalues are
\begin{eqnarray*}
\label{eigenv_sc}
\lambda_{kl}=2|\breve n|B(k+l+1), \, \, k,l=0,1,\dots.
\end{eqnarray*}
The amplitudes $f(p_1)$ and $g(p_4)$ have to provide square integrability of the eigenfunctions in $x_1$ and $x_4$. In order to satisfy condition (\ref{current_cont_sc}) one has to restrict the amplitude $f(p_1)$ additionally. 
The integral current through the domain wall is continuous if both $f$ and $H_k$ are  odd or even functions simultaneously under the combined change $p_1\to-p_1$ and $x_2\to-x_2$
\begin{eqnarray}
 \label{cont_cond_sc}
f(-p_1)H_k(-z)=f(p_1)H_k(z).
\end{eqnarray}
This property also guarantees the absence of the gauge specific contribution to the action  related to the derivative of $H_3$ in Eqs.(\ref{sing_gauge},\ref{surf_gauge}). 

A combination of (\ref{cont_cond_sc}) and (\ref{conditions}) obviously leads to the relation
\begin{eqnarray*}
 \label{relations}
\phi_{k}^{(\pm)}(x_1,x_2)=\phi_{k}^{(\pm)}(-x_1,-x_2),
\end{eqnarray*}
where $(\pm)$ denotes the duality of the background field for a given $x_1$. This identity allows one to show that the eigenfunctions 
\begin{equation*}
\Phi_{kl}(x) = \left\{ 
\begin{array}{l}
\Phi^{(+)}_{kl}(x), \, \, x_1\in L_+\\
\Phi^{(-)}_{kl}(x), \, \, x_1\in L_-
\end{array} 
\right., \, k,l=0,1\dots
\end{equation*}
form a complete orthogonal set in the space of square integrable functions which are even with respect to simultaneous reflection $x_1\to -x_1$ and $x_2\to-x_2$.

The eigenfunctions are of the bound state type with the purely discrete spectrum. Field fluctuations of this type can be seen as confined. It should be noted that the eigenvalues coincide with those for the purely homogeneous (anti-)self-dual Abelian field. In this sense, the domain wall defect does not destroy dynamical confinement of color charged fields. The eigenfunctions are restricted by the correlated evenness condition (\ref{cont_cond_sc}), while in the case of the  homogeneous field the properties of the amplitude $f(p_1)$ and the polynomial $H_k$ are mutually independent.

\subsection{Color charged quasiparticles on the wall}

Let us now consider the eigenvalue problem on the domain wall, i.e. for the region $x_1=0$. On the wall the chromomagnetic and chromoelectric fields are orthogonal to each other (see Fig.\ref{Fig:single_kink}).
In conformity with (\ref{current_cont_sc}) the absence of the charged current off the infinitely thin domain wall requires
\begin{eqnarray*}
 \partial_1\Phi|_{x_1=0}=0,
\end{eqnarray*}
and the eigenvalue problem on the wall takes the form
\begin{eqnarray}
\label{covder_wall}
\left[- \partial_2^2 -\partial_3^2 +\breve n^2 B^2 x_3^2 + (i\partial_4-\breve n Bx_3)^2\right]\Phi=\lambda\Phi
\nonumber
\end{eqnarray}
with the solution
\begin{eqnarray}
\label{on_wall_sol_sc}
\Phi_k(x_2,x_3,x_4)&=&e^{ip_2x_2+ip_4x_4}e^{-\frac{|\breve n|B}{\sqrt{2}}\left(x_3-\frac{p_4}{2|\breve n|B}\right)^2}
\nonumber\\
&\times&
H_k\left[\sqrt{\sqrt{2}|\breve n|B}\left(x_3-\frac{p_4}{2|\breve n|B}\right)\right],
\nonumber\\
\lambda_k(p^2_2,p_4^2)&=&\sqrt{2}|\breve n|B(2k+1)+\frac{p_4^2}{2}+\frac{p_2^2}{2},
\nonumber\\
 &&k=0,1,2,\dots
\nonumber
\end{eqnarray}
The spectrum of the eigenmodes on the wall is continuous, it depends on the momentum $p_2$ longitudinal to the chromomagnetic field and Euclidean energy $p_4$, the corresponding eigenfunctions are oscillating in $x_2$ and $x_4$. In the direction $x_3$ transverse to the chromomagnetic field  the eigenfunctions are bounded and the eigenvalues display the Landau level structure.  The continuation $p_4^2=-p_0^2$ leads to the  dispersion relation 
 \begin{equation*}
p_0^2=p_2^2+\mu^2_k,\quad \mu^2_k=2\sqrt2(2k+1)|\breve n|B,\, \, k=0,1,2,\dots .  
 \end{equation*}
This can be treated as the lack of confinement - the color charged quasiparticles with masses $\mu_k$ and momentum $\mathbf{p}$ parallel to the chromomagnetic field $\mathbf{H}$ can be excited on the wall.

The case of the planar domain wall configuration (two infinite parts of the space-time separated by a three-dimentional hypersurface like in Fig.\ref{Fig:single_kink}) is rather artificial. Its weight in the whole ensemble of the gluon field configurations with the constant  scalar condensate $\langle g^2F_{\mu\nu}F_{\mu\nu}\rangle$ and the lumpy structured distribution of the topological charge density $\langle g^2\tilde F_{\mu\nu}F_{\mu\nu}\rangle$ is negligible. The entropy-energy balance  implies that the typical configuration should be highly disordered (see Fig.\ref{Fig:kink_network}).  Moreover, in the case of the planar domain wall the eigenvalue problem for the square integrable vector gauge fields
\begin{eqnarray}
\label{vector}
\left[-D^2\delta_{\mu\nu}+2i\breve F_{\mu\nu}\right]Q_\nu=\lambda Q_\mu
\end{eqnarray}
 leads to the negative eigenvalues and corresponding tachyonic modes on the wall where $\tilde F_{\mu\nu}F_{\mu\nu}=0$.  This is a well-known instability of the Nielsen-Olesen type~\cite{Nielsen:1978rm}. The presence of the tachyonic mode is due to the three infinite dimensions of the planar domain wall hypersurface. One can expect that  finite size of boundaries between lumps in the typical kink network configuration, Fig.\ref{Fig:kink_network}, removes the tachyonic modes.
This is manifestly exemplified in the next section where the color charged field eigenvalues and modes are studied for thick cylindrical domain wall junction. The relatively stable defect of this type can occur in the ensemble of confining gluon fields due to the influence of the strong electromagnetic fields on the QCD vacuum structure.    

\section{The spectrum of color charged quasiparticles trapped in a thick domain wall junction}
\label{trap}

\subsection{Heavy ion collisions: the strong electromagnetic field as a trigger for deconfinement}

It has been observed that  the strong electromagnetic fields  generated in relativistic heavy ion collisions  can play the role of a trigger for  deconfinement~\cite{NG2011-1}.  The mechanism discussed 
in~\cite{NG2011-1} is as follows. The electric $\mathbf E_{\rm el}$ and magnetic $\mathbf H_{\rm el}$ fields are practically orthogonal to each other~\cite{toneev,Skokov:2009qp}: $\mathbf E_{\rm el}\mathbf H_{\rm el}\approx 0 $. For this configuration of the external electromagnetic field the one-loop
quark contribution to the QCD effective potential for the
homogeneous Abelian gluon fields is minimal for the  chromoelectric and chromomagnetic fields directed along the electric and magnetic fields respectively.  The orthogonal chromo-fields are not confining: color charged quasiparticles can  move along the chromomagnetic field. It has been noted also that this mechanism  assumes the strong azimuthal anisotropy in momentum distribution of color charged quasiparticles.   Deconfined quarks as well as gluons will move preferably along
the direction of the magnetic field but this will happen due to
the gluon field configuration even
after switching the electromagnetic field off.

A detailed and systematic analytical one-loop calculation of the QCD effective potential for the pure chromomagnetic field was performed recently in \cite{Ozaki:2013sfa} and confirmed the result that the chromomagnetic field prefers to be parallel (or anti-parallel) to the external magnetic field. Another important source of verification of the basic observations of paper~\cite{NG2011-1} is due to the recent  Lattice QCD studies of the response of the QCD vacuum to external electromagnetic fields~\cite{D'Elia:2012zw,Bali:2013esa,Bali:2013owa,Bonati:2013qra}. 

In particular, in qualitative agreement with~\cite{NG2011-1}  Lattice QCD study~\cite{Bali:2013esa}  has demonstrated that  in the presence of external magnetic field the gluonic 
action develops an anisotropy: the chromomagnetic
field parallel to the external field is enhanced, while the chromo-electric field in this direction is suppressed. The results of \cite{Bali:2013owa} indicated that the magnetic field  can affect the azimuthal structure of the expansion of the system during heavy ion collisions.

\begin{figure}[h!]
 \centerline{\includegraphics[width=38mm,angle=0]{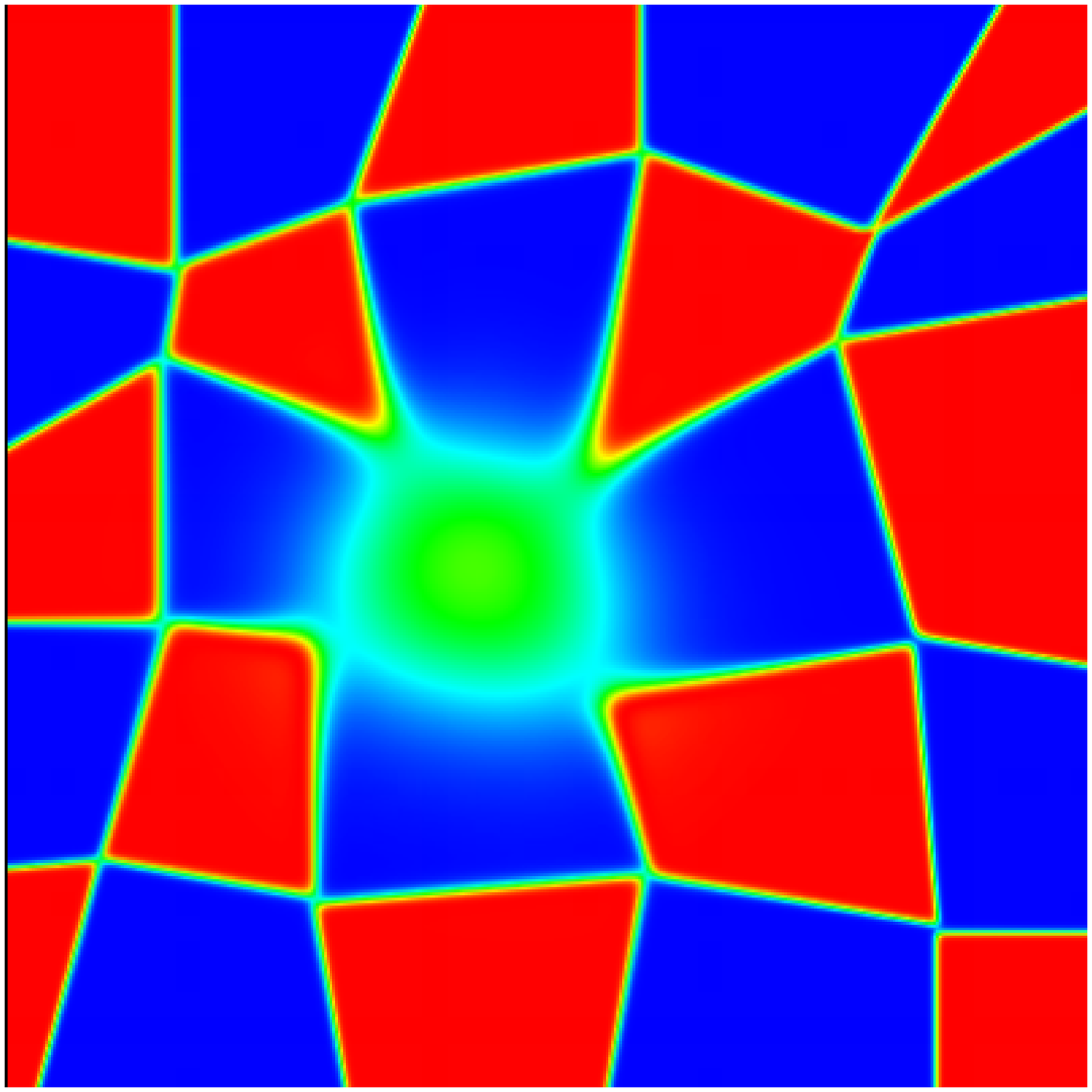}\hspace*{2mm}\includegraphics[width=38mm,angle=0]{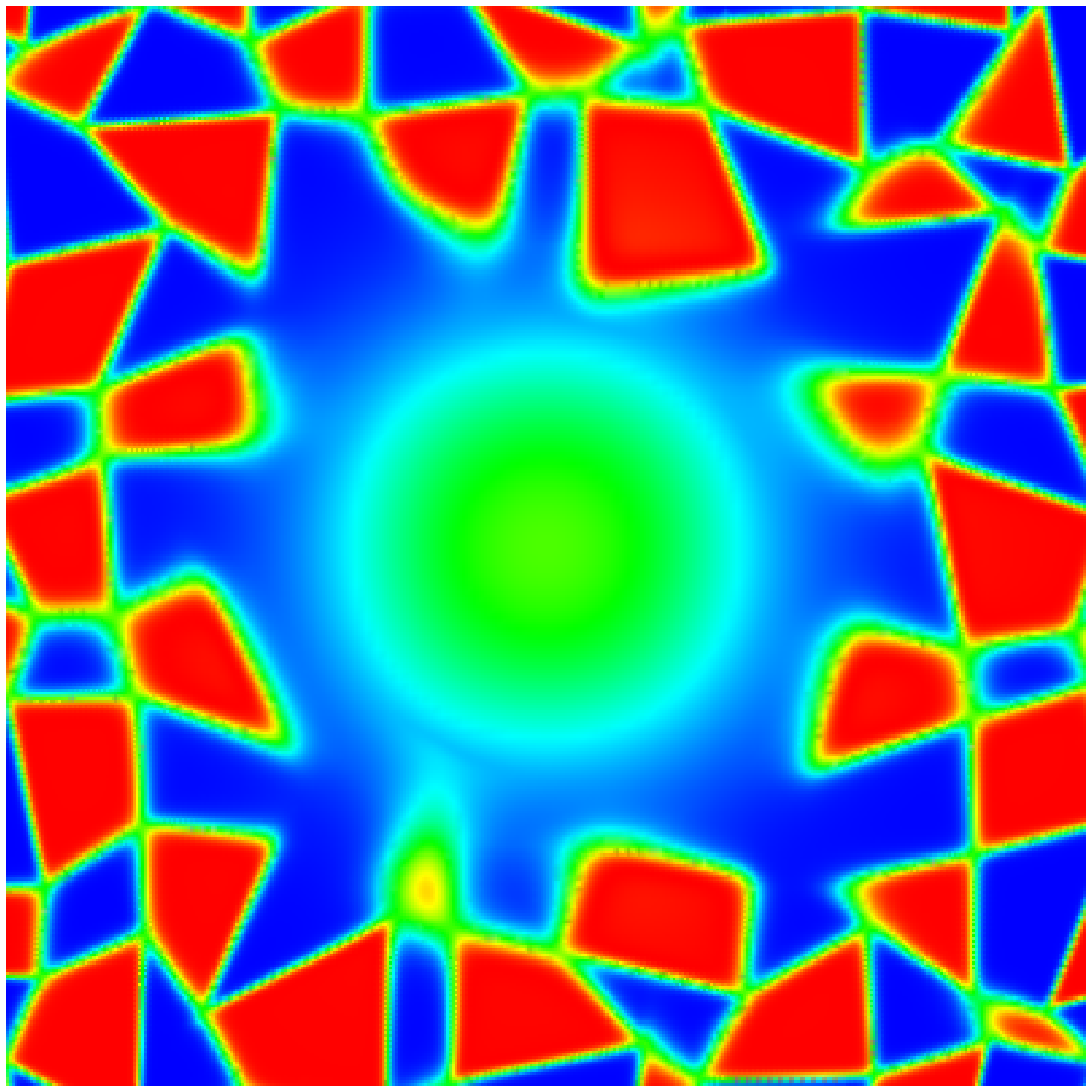}}
    \caption{ Examples of two-dimensional slice of the cylindrical thick domain wall junctions. The correspondence of colors  is the same as in Fig.\ref{Fig:layered}. Blue and red regions represent self-dual and anti-self-dual lumps. Confinement is lost in the green region where 
 $g^2\tilde F_{\mu\nu}(x)F_{\mu\nu}(x)=0$. The scalar condensate density $g^2F_{\mu\nu}(x)F_{\mu\nu}(x)$ is nonzero and homogeneous everywhere.}
 \label{Fig:chromo_bag}
\end{figure}

 Within the context of the confining domain wall network these observations mean that a flash of the strong electromagnetic field during heavy ion collisions produces a kind of defect in the form of the thick domain wall junction  in the confining gluon background exactly in the region where collision occurs (see Fig.\ref{Fig:chromo_bag}). The electromagnetic flash can act as one of the preconditions for  conversion of the high energy density and baryon density  to the
thermodynamics of color charged degrees of freedom. \\

\subsection{ Cylindrical trap}

\subsubsection{Scalar field eigenmodes}

Since topological charge density is zero in the interior of the trap ($g^2\tilde F_{\mu\nu}(x)F_{\mu\nu}(x)=0$) there exists a specific reference frame where one can use  the pure chromomagnetic field for description of the gluon background inside the trap. For simplicity we take  cylindrical geometry of the trap and study the properties of scalar and vector (gluon) color charged field eigenmodes.  Extension of the present consideration to more realistic form of the trap  is straightforward.
 
Consider the eigenvalue problem for the massless scalar field $\Phi^a$
\begin{eqnarray}
\label{cylinder_s}
-\left(\partial_\mu-i\breve B_\mu\right)^2\Phi(x)=\lambda^2\Phi(x)
\end{eqnarray}
in the cylindrical region
\begin{eqnarray*}
x\in\mathcal{T}=\left\{x_1^2+x_2^2<R^2, \ (x_3,x_4)\in \mathrm{R^2} \right\}
\end{eqnarray*}
with the homogeneous Dirichlet condition at the boundary
\begin{eqnarray} 
\label{HDC}
&&\Phi(x)=0, \  x\in \partial\mathcal{T} 
\\
&&\partial\mathcal{T} =\left\{x_1^2+x_2^2=R^2, \ (x_3,x_4)\in \mathrm{R^2}  \right\}.
\nonumber
\end{eqnarray}
Here $\breve B_\mu$ stays for adjoiunt representaion of the homogeneous chromomagnetic field $H^a_i=\delta_{i3}n^a H$ with the vector
potential taken in the symmetric gauge
\begin{gather}
\breve B_\mu=-\frac{1}{2} \breve n B_{\mu\nu}x_\nu, 
\label{purechrmag}\\
\breve B_4=\breve B_3=0, \ B_{12}=-B_{21}=H,
\nonumber\\
\breve n=T_3\cos(\xi)+T_8\sin(\xi).
\nonumber
\end{gather}
\begin{widetext}
The eigenvalues of the matrix $\breve n$ are
\begin{eqnarray}
\label{n_diag}
\breve v=\mathrm{diag}\left[\cos\left(\xi\right),-\cos\left(\xi\right),0,\cos\left(\xi-\frac{\pi}{3}\right),
-\cos\left(\xi-\frac{\pi}{3}\right),
\cos\left(\xi+\frac{\pi}{3}\right),-\cos\left(\xi+\frac{\pi}{3}\right),0\right].
\label{vadj}
\end{eqnarray}
\end{widetext}
For any value of the angle $\xi$ there are two zero eigenvalues  $\breve{v}_3=\breve{v}_8=0$.  Two
additional zero elements occur in $\breve{v}$ if the angle takes values $\xi_k$ (see Eq.~\eqref{HLxik}) minimizing the effective potential \eqref{ueff} and corresponding to the boundaries of the Weyl chambers. By inspection one can check that  nonzero eigenvalues $\breve{v}$
take  values $\pm v$ with $v=\sqrt{3}/2$.  Below we use  notation
\begin{eqnarray*}
\nonumber
\breve v^a=v\kappa^a.
\end{eqnarray*}
For example, if $\xi=\xi_0=\pi/6$ then the nonzero values of $v^a$ correspond to $a=1,2,4,5$ and 
\begin{eqnarray*}
\nonumber
\kappa_1=1, \kappa_2=-1, \kappa_4=1, \  \kappa_5=-1.
\end{eqnarray*}
 It has to be noted that the  effective Lagrangian \eqref{ueff} leads to the kink configuration (for details see \cite{NG2011})
\begin{eqnarray*}
 \xi_{k}(x_i) =\frac{1}{3} \arctan\left[\sinh(m_\xi x_i)\right]+\frac{\pi k}{3}, \ k=0,\dots,5,
\end{eqnarray*}
interpolating between boundaries $\xi_k$ and $\xi_{k+1}$ of the $k$-th Weyl chamber. Superposition of these "color" domain walls can be arranged in a complete analogy with the "duality" domain walls. The only new feature of the "color" domains is that there are six different types interrelated by the Weyl reflections instead of two types as in the case of duality domains. 

Solution of the problem \eqref{cylinder_s} is straightforward. We give it below just for completeness.

It is convenient to introduce dimensionless variables using the strength of the chromomagnetic field as a basic scale. Below all quantities are assumed to be measured in terms of this scale, for instance
\begin{equation*}
\sqrt{H}x_\mu \equiv x_\mu,\quad \frac{\lambda}{\sqrt{H}}\equiv \lambda.
\end{equation*}

 After diagonalization  with respect to color indices and transformation to the cylindrical coordinates Eq.~\eqref{cylinder_s}  takes the form 
\begin{eqnarray}
\label{HDC1}
-\left[\partial_4^2+\partial_3^2
+\frac{\partial^2}{\partial r^2}+\frac{1}{r}\frac{\partial}{\partial  r}+\frac{1}{r^2}\frac{\partial^2}{\partial \vartheta^2}
-i\kappa^a v \frac{\partial}{\partial \vartheta}
\right.
\nonumber\\
\left.
 -\frac{1}{4} v^2r^2\right]\Phi^a=\lambda^2\Phi^a,\ \ \
\end{eqnarray}
where it has been used that
\begin{eqnarray*}
&&x_1=r\cos\vartheta,\ x_2=r\sin\vartheta,\\
\\
&&\frac{\partial}{\partial x_1}=\cos\vartheta\frac{\partial}{\partial r}-\frac{\sin\vartheta}{r}\frac{\partial}{\partial \vartheta},
\\
&&\frac{\partial}{\partial x_2}=\sin\vartheta\frac{\partial}{\partial r}+\frac{\cos\vartheta}{r}\frac{\partial}{\partial \vartheta},
\\
&&\partial_1^2+\partial_2^2=\frac{\partial^2}{\partial r^2}+\frac{1}{r}\frac{\partial}{\partial  r}+\frac{1}{r^2}\frac{\partial^2}{\partial \vartheta^2}.
\end{eqnarray*}

The variables in Eq.~\eqref{HDC1} are separated by substitution 
\begin{eqnarray*}
\Phi^a=\phi^a(r)e^{il\vartheta}\exp\left(ip_3x_3+ip_4x_4\right).
\end{eqnarray*}
Periodicity of the solution in angle  $\vartheta\in[0,2\pi]$ requires integer values of parameter $l$. 

The radial part $\phi(r)$ should satisfy equation
\begin{equation}
\label{HDCR}
-\left[\frac{\partial^2}{\partial r^2}+\frac{1}{r}\frac{\partial}{\partial  r}-\frac{1}{r^2}\left(\frac{1}{2}\breve v r^2-l\right)^2
\right]\phi=\mu^2\phi,
\end{equation}
where $\mu$ is related to the original eigenvalue $\lambda$,
\begin{eqnarray*}
\lambda^2=p_4^2+p_3^2+\mu^2.
\end{eqnarray*} 

By means of  the substitution
\begin{equation*}
\phi=r^le^{-\frac{1}{4}\breve{v}r^2}\chi,   
\end{equation*}
one arrives at the Kummer equation ($z=\breve{v}r^2/2$)
\begin{equation}
\left[z\frac{d^2}{dz^2}+(l+1-z)\frac{d}{dz}-\frac{\breve{v}-\mu^2}{2\breve{v}}\right]\chi=0.
\end{equation}
The complete solution can be chosen in the form 
\begin{widetext}
\begin{equation*}
\chi(z)=C_1M\left(\frac{\breve{v}-\mu^2}{2\breve{v}},1+l,z\right)+C_2z^{-l}M\left(\frac{\breve{v}-\mu^2}{2\breve{v}}-l,1-l,z\right),
\end{equation*}
where $M(a,b,z)$ is Kummer function.
General solution of equation \eqref{HDCR} takes the form
\begin{equation*}
\phi_{l}(r)=
e^{-\frac{1}{4}\breve{v}r^2}\left[C_1r^lM\left(\frac{\breve{v}-\mu^2}{2\breve{v}},1+l,\frac12\breve{v}r^2\right)
+C_2r^{-l}M\left(\frac{\breve{v}-\mu^2}{2\breve{v}}-l,1-l,\frac12\breve{v}r^2\right)\right]
\end{equation*}
The first term is regular at $r=0$ provided $l\geqslant 0$ while the second one is well-defined for $l\leqslant 0$. Therefore, the solution  regular inside the cylinder  is
\begin{eqnarray}
\phi_{al}&=&e^{-\frac{1}{4}\breve{v}_ar^2}r^lM\left(\frac{\breve{v}_a-\mu^2}{2\breve{v}_a},1+l,\frac12\breve{v}_ar^2\right),  \ \  l\geqslant 0,
\\
\phi_{al}&=&e^{-\frac{1}{4}\breve{v}_ar^2}r^{-l}M\left(\frac{\breve{v}_a-\mu^2}{2\breve{v}_a}-l,1-l,\frac12\breve{v}_ar^2\right), \ \ l< 0,
\label{scalarphi}
\end{eqnarray}
where the color index $a$ has been explicitly indicated. The color matrix elements $\breve{v}_a$ can be negative. In this case one has to apply Kummer transformation~\cite{AS}
\begin{equation*}
M(a,b,z)=e^zM(b-a,b,-z).
\end{equation*}
\end{widetext}

Dirichlet boundary condition \eqref{HDC} defines the infinite discrete set of eigenvalues as the solutions $\mu^2_{alk}$ ($k=0,1\dots\infty$)  of the equations
\begin{eqnarray}
\label{eigenvaluesscalar1}
M\left(\frac{\hat{v}_a-\mu^2}{2\hat{v}_a},1+l,\frac12\hat{v}_aR^2\right)=0,\quad l\geqslant 0,
\\
M\left(\frac{\hat{v}_a-\mu^2}{2\hat{v}_a}-l,1-l,\frac12\hat{v}_aR^2\right)=0,\quad l< 0.
\label{eigenvaluesscalar2}
\end{eqnarray}
If $\mu^2_{alk}$ satisfies equation \eqref{eigenvaluesscalar1}, than $\tilde\mu^2_{alk}=\mu^2_{alk}-2\hat{v}_al$ is a solution of \eqref{eigenvaluesscalar2}.

\begin{figure}[h!]
 \centerline{\includegraphics[width=75mm,angle=0]{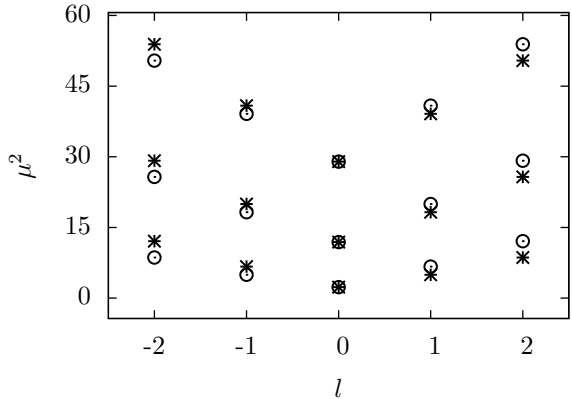}}
    \caption{Eigenvalues $\mu^2_{alk}$ for the scalar field problem, $l=-2,-1,0,1,2$ and $k=0,1,2$, for $\sqrt{H}R=1.6$. 
Eigenvalues are denoted by asterisks in the case of positive $v_a$ and 
by circles in the case of negative $v_a$. }
 \label{Fig:Eigen_set1}
\end{figure}

Finally the complete orthogonal set of eigenfunctions for the problem \eqref{cylinder_s} and \eqref{HDC} reads
\begin{eqnarray*}
&&\Phi_{alk}(p_3, p_4|r,\vartheta,x_3,x_4)=e^{ip_3x_3+ip_4x_4}
e^{il\vartheta}\phi_{alk}(r),
\nonumber
\\
&&\lambda_{alk}^2=p_4^2+p_3^2+\mu_{akl}^2,
\label{eigenv_eucl}
\\
&& k=0,1,\dots,\infty, \ \ l=-\infty\dots\infty,
\nonumber
\end{eqnarray*}
where functions $\phi_{alk}$ are defined by \eqref{scalarphi} with  $\mu^2=\mu_{akl}^2$ solving the boundary condition \eqref{eigenvaluesscalar1}.
Unlike Landau levels in the infinite space the eigenvalues $\mu_{akl}^2$ are not equidistant in $k$ and non-degenerate in $l$ as it is illustrated in Fig.\ref{Fig:Eigen_set1}). The dependence of several low-lying eigenvalues $\mu_{akl}^2$ on the dimensionless size parameter $\sqrt{H}R$
is shown in Fig.\ref{Fig:Eigen_flow}.

\subsubsection{Vector field eigenmodes}

For pure chromomagnetic field \eqref{purechrmag} the adjoint representation vector field 
Eq.~\eqref{vector} takes the form 
\begin{eqnarray}
\label{vector1}
\left[-\breve D^2\delta_{\mu\nu}+2i\breve n B_{\mu\nu}\right]Q_\nu=\lambda Q_\mu,
\end{eqnarray}
and the boundary conditions are
\begin{eqnarray} 
&&\breve n Q_{\mu}(x)=0, \  x\in \partial\mathcal{T} 
\nonumber\\
&&\partial\mathcal{T} =\left\{x_1^2+x_2^2=R^2, \ (x_3,x_4)\in \mathrm{R^2}  \right\}
\label{HDCV}
\end{eqnarray}
In terms of the eigenvectors $\breve{Q}_\mu^a$ of matrices $B_{\mu\nu}$ and $\breve n$
Eqs.~\eqref{vector1} and \eqref{HDCV} take the form
\begin{eqnarray}
\label{vector2}
\left[-\breve D^2+2s_{\mu}\breve v H\right]^a\breve{Q}^{a}_\mu 
=\lambda_{a\mu}\breve{Q}^{a}_\mu,
\\\nonumber
\breve v \breve{Q}_{\mu}(x)=0, \  x\in \partial\mathcal{T}.
\end{eqnarray}

Omitting obvious well-known details we just note that equation \eqref{vector2} describes sixteen charged with respect to $\breve n$ spin-color polarizations  of the gluon fluctuations with $(s_1=1, s_2=-1, s_3=s_4=0)$ and $\breve v^a\not=0$ 
as well as sixteen  "color neutral" with respect to $\breve{n}$ modes 
\begin{eqnarray*}
\label{vector4}
&&-\partial^2\breve{Q}_\mu^{(0)}=p^2 \breve{Q}_\mu^{(0)}.
\label{Q0}
\end{eqnarray*}
Neutral mode $\breve{Q}_\mu^{(0)}$ is a zero mode of $\breve n$, and it is insensitive to the boundary condition \eqref{HDCV}. We shall briefly discuss the possible role of the  neutral modes in the last section.

\begin{figure}[h!]
 \centerline{\includegraphics[width=75mm,angle=0]{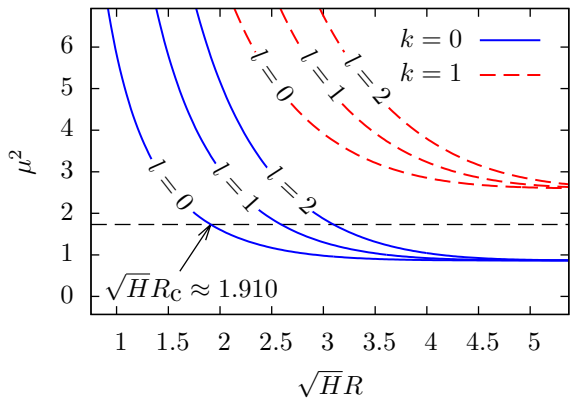}}
    \caption{The lowest eigenvalues corresponding to positive color orientation $\kappa^a=1$ as functions of $\sqrt{H}R$. The critical radius $R_{\rm c}$ corresponds to $\mu_{a00}^2=2v=\sqrt{3}$. For large $\sqrt{H}R$ eigenvalues  approach correct Landau levels, the degeneracy in $l$ is restored.}
 \label{Fig:Eigen_flow}
\end{figure}

Equations for the color charged modes have the same form as the scalar field equation in the previous subsection. The only essential difference is that the eigenvalues $\lambda_{alk\nu}$ for nonzero $v^a$ have an addition
$\pm 2vH$ to the eigenvalues $\mu^2_{akl}$ of the scalar case:
\begin{eqnarray*}
 &&\lambda^2_{alk\nu}=p_4^2+p_3^2+\mu_{alk}^2+ 2s_\nu\kappa_a v,
\label{eigenv_eucl_pm}
\\
&& k=0,1,\dots,\infty, \ \ l\in Z,
\nonumber\\
&& s_1=1, \ s_2=-1, \ s_3=s_4=0, \ \kappa_a=\pm 1.
\nonumber
\end{eqnarray*}
where $\mu^2_{akl}$ are the same as in the scalar case. If we were considering the square integrable solutions in $R^4$ then the lowest mode $\lambda^2_{a00\nu}$ with $s_\nu\kappa_a=-1$
 would be tachyonic.   In the finite trap the lowest eigenvalue is
\begin{eqnarray*}
\label{lambda00}
\lambda^2_{a00\nu}=p_4^2+p_3^2+\mu_{a00}^2-2v, \  s_\nu\kappa_a=-1.
\end{eqnarray*}
The dependence of $\mu_{a00}^2$ on  dimensionless size parameter $\sqrt{H}R$ is strongly nonlinear. 
Few lowest eigenvalues $\mu_{akl}^2$ as functions of $\sqrt{H}R$ are shown in Fig.~\ref{Fig:Eigen_flow}. 
One concludes that if the dimensionless size $\sqrt{H}R$ of the trap is sufficiently small 
\begin{eqnarray}
\label{dmlsc}
\sqrt{H}R<\sqrt{H}R_{\rm c}\approx 1.91,
\end{eqnarray}
  then there are no unstable tachyonic modes in the spectrum of color charged vector fields.
  
To estimate the critical size one may use the mean phenomenological value of the gluon condensate (gauge coupling constant $g$ is included into the field strength tensor) 
\begin{eqnarray*}
\langle F^a_{\mu\nu}F^{a\mu\nu}\rangle = 2H^2\approx 0.5{\rm GeV}^4.
\end{eqnarray*} 
Equation~\eqref{dmlsc} leads to the critical radius
\begin{eqnarray}
R_{\rm c}\approx 0.51 \ {\rm fm} \ (2R_{\rm c}\approx 1 \ {\rm fm}).
\label{rc}
\end{eqnarray}
 Thus the tachyonic mode is absent if the diameter of the cylindrical trap is less or equal to  $1 \ {\rm fm}$.

\subsubsection{Quark field eigen modes}

In this subsection we address the eigenvalue problem for  Dirac operator in the cylindrical region  in the presence of chromomagnetic background field
(\ref{purechrmag}) 
\begin{eqnarray}
&&\!\not\!\!D\psi(x)=\lambda\psi(x),
\label{dirac}\\
&&D_\mu=\partial_\mu+\frac{i}{2}\hat n B_{\mu\nu}x_\nu,
\nonumber\\
&&\hat n= t_3\cos\xi +t_8\sin\xi
\label{nfund}\\
&&=\frac{1}{2}{\rm diag}\left(\cos\xi+\frac{\sin\xi}{\sqrt{3}}, -\cos\xi+\frac{\sin\xi}{\sqrt{3}}, -\frac{2\sin\xi}{\sqrt{3}}\right).
\nonumber
\end{eqnarray}
Euclidean Dirac matrices are taken in the anti-hermitian representation. 

The angle $\xi$ is assumed to take one of the vacuum values $\xi_k$, and according to Eq.~\eqref{HLxik} the following forms of the matrix
$\hat n$ can occur 
\begin{eqnarray}
&&\hat n=\left\{ \pm \frac{1}{\sqrt{3}}{\rm diag}\left(1, -\frac{1}{2}, -\frac{1}{2}\right),
\right.
\label{HLxikfund}\\
&&\left.
\pm \frac{1}{\sqrt{3}}{\rm diag}\left(\frac{1}{2}, \frac{1}{2},-1\right), \pm \frac{1}{\sqrt{3}}{\rm diag}\left(-\frac{1}{2},1, -\frac{1}{2}\right) \right\}.
\nonumber
\end{eqnarray}
Below we use notation
\begin{eqnarray*}
\hat{n}_{ij}=\delta_{ij}\hat{u}_j.
\nonumber\\
\end{eqnarray*}
The boundary conditions are
\begin{eqnarray}
i\!\not\!\eta(x)e^{i\theta\gamma_5}\hat n\psi(x)  =  \hat n\psi(x), \ x\in\partial\mathcal{T},
\nonumber\\
\bar\psi(x)e^{i\theta\gamma_5} \hat n i \! \!\not\!\eta(x)  =  -\bar\psi(x) \hat n, 
\ x\in\partial\mathcal{T},
\label{bagbc}
\end{eqnarray}
where $\eta_\mu$ is a unit vector normal to the cylinder surface $\partial\mathcal{T}$, see Eq.~\eqref{HDCV}.  
These are simply the bag boundary conditions. This choice appears to be rather natural. Indeed, inside the thick domain wall junction one expects an existence of the color charged quasiparticles (quarks) being the carriers of the color current, but outside the junction gluon configurations are confining (see Fig.~\ref{Fig:chromo_bag})) and the   current has to vanish at the boundary.  
Unlike the adjoint representation of color matrix \eqref{vadj} the matrix $\hat n$ in fundamental representation \eqref{nfund}
has no zero eigenvalues for any value of the angle $\xi$ corresponding to the boundaries of Weyl chambers, see Eq.~\eqref{HLxikfund}. Boundary condition \eqref{bagbc} restricts all three color components of the quark field.   

Substitution
\begin{equation}\label{psirelation}
\psi=\left(\not{\hspace*{-0.3em}D}+\lambda\right)\varphi
\end{equation}
leads to the equation 
\begin{eqnarray}
\label{squareD}
-\left(D^2+\hat{u}H\Sigma_3\right)\varphi=\lambda^2\varphi,
\end{eqnarray}
where it has been used that in the pure chromomagnetic field \eqref{purechrmag}
\begin{eqnarray*}
 \frac12\sigma_{\mu\nu}\hat{B}_{\mu\nu}=\Sigma_3H\hat{u},
 \ \Sigma_i=\frac12\varepsilon_{ijk}\sigma_{jk}.
\end{eqnarray*}
Equation \eqref{squareD} is essentially the same as \eqref{cylinder_s}. Its solution in cylindrical coordinates ($2\pi$-periodic in $\vartheta$ and regular at $r=0$) is given by four independent components $\varphi_l^\alpha$ ($\alpha=1,\dots,4$, $l\in Z$):
\begin{eqnarray*}
\varphi_l^\alpha=e^{-ip_3x_3-ip_4x_4}e^{il\vartheta}\phi_l^\alpha(r)
\end{eqnarray*}
with
\begin{equation*}
\label{philp}
\phi_{l}^\alpha=e^{-\frac14\hat{u}r^2}r^{l}
M\left(\frac{1+s_\alpha}{2}-\frac{\mu^2}{2\hat{u}},1+l,\frac{\hat{u}r^2}{2}\right)
\end{equation*}
for the case $l\geqslant 0$ and
\begin{equation*}
\label{philn*}
\phi_{l}^{\alpha}=e^{-\frac14\hat{u}r^2}r^{-l}
M\left(\frac{1+s_{\alpha}}{2}-\frac{\mu^2}{2\hat{u}}-l,1-l,\frac{\hat{u}r^2}{2}\right)
\end{equation*}
for  $l<0$. Here 
$$s_\alpha=(-1)^\alpha, \ \alpha=1,\dots,4$$
denotes the sign of the quark spin projection on the direction of chromomagnetic field, and therefore
\begin{eqnarray*}
\label{philarrows}
\phi_l^3=\phi_l^1=\Phi_l^{\uparrow\uparrow}(r), \ \phi_l^4=\phi_l^2=\Phi_l^{\uparrow\downarrow}(r).
\end{eqnarray*}

The variable $\mu$ is related to the Dirac eigenvalues as 
\begin{equation*}
\mu^2=\lambda^2-p_3^2-p_4^2.
\label{mu}
\end{equation*}

\begin{widetext}
Finally the Dirac operator eigenfunction $\psi$ can be obtained by means of relation \eqref{psirelation} with
\begin{equation}
\not{\hspace*{-0.3em}D}+\lambda=\left(
\begin{array}{cccc}
\lambda&0&i\partial_4+\partial_3&D_1-iD_2\\
0&\lambda&D_1+iD_2&i\partial_4-\partial_3\\
i\partial_4-\partial_3&-D_1+iD_2&\lambda&0\\
-D_1-iD_2&i\partial_4+\partial_3&0&\lambda
\end{array}
\right),
\nonumber
\end{equation}
\end{widetext}
where
\begin{gather*}
D_1+iD_2=e^{i\vartheta}\left(\frac{\partial}{\partial r}+\frac{i}{r}\frac{\partial}{\partial \vartheta}+\frac12\hat{u}r\right),\displaybreak[0]
\\
D_1-iD_2=e^{-i\vartheta}\left(\frac{\partial}{\partial r}-\frac{i}{r}\frac{\partial}{\partial \vartheta}-\frac12\hat{u}r\right).
\end{gather*}
Four solutions are for $l\geqslant 0$
\begin{equation*}
\psi_l^{(1)}=e^{-ip_3x_3-ip_4x_4}\left(
\begin{array}{c}
\lambda \Phi_l^{\uparrow\uparrow}(r)e^{il\vartheta}\\
0\\
(p_4+ip_3)\Phi_l^{\uparrow\uparrow}(r)e^{il\vartheta}\\
\frac{\mu^2}{2(l+1)}\Phi_{l+1}^{\uparrow\downarrow}(r)e^{i(l+1)\vartheta}
\end{array}
\right)
\end{equation*}
\begin{equation*}
\psi_l^{(2)}=e^{-ip_3x_3-ip_4x_4}\left(
\begin{array}{c}
0\\
\lambda \Phi_{l+1}^{\uparrow\downarrow}(r)e^{i(l+1)\vartheta}\\
-2(l+1)\Phi_l^{\uparrow\uparrow}(r)e^{il\vartheta}\\
(p_4-ip_3)\Phi_{l+1}^{\uparrow\downarrow}(r)e^{i(l+1)\vartheta}
\end{array}
\right)
\end{equation*}
\begin{equation*}
\psi^{(3)}_l=e^{-ip_3x_3-ip_4x_4}\left(
\begin{array}{c}
(p_4-ip_3)\Phi_{l}^{\uparrow\uparrow}(r)e^{il\vartheta}\\
-\frac{\mu^2}{2(1+l)}\Phi_{l+1}^{\uparrow\downarrow}(r)e^{i(l+1)\vartheta}\\
\lambda \Phi_l^{\uparrow\uparrow}(r)e^{il\vartheta}\\
0
\end{array}
\right)
\end{equation*}
\begin{equation*}
\psi^{(4)}_l=e^{-ip_3x_3-ip_4x_4}\left(
\begin{array}{c}
2(l+1)\Phi_l^{\uparrow\uparrow}(r)e^{il\vartheta}\\
(p_4+ip_3)\Phi_{l+1}^{\uparrow\downarrow}(r)e^{i(l+1)\vartheta}\\
0\\
\lambda \Phi_{l+1}^{\uparrow\downarrow}(r)e^{i(l+1)\vartheta}\\
\end{array}
\right),
\end{equation*}
and for $l<0$
\begin{equation*}
\psi_l^{(1)}=e^{-ip_3x_3-ip_4x_4}\left(
\begin{array}{c}
\lambda \Phi_l^{\uparrow\uparrow}(r)e^{il\vartheta}\\
0\\
(p_4+ip_3)\Phi_l^{\uparrow\uparrow}(r)e^{il\vartheta}\\
2l\Phi_{l+1}^{\uparrow\downarrow}(r)e^{i(l+1)\vartheta}
\end{array}
\right)
\end{equation*}
\begin{equation*}
\psi_l^{(2)}=e^{-ip_3x_3-ip_4x_4}\left(
\begin{array}{c}
0\\
\lambda \Phi_{l+1}^{\uparrow\downarrow}(r)e^{i(l+1)\vartheta}\\
-\frac{\mu^2}{2l}\Phi_l^{\uparrow\uparrow}(r)e^{il\vartheta}\\
(p_4-ip_3)\Phi_{l+1}^{\uparrow\downarrow}(r)e^{i(l+1)\vartheta}
\end{array}
\right)
\end{equation*}
\begin{equation*}
\psi^{(3)}_l=e^{-ip_3x_3-ip_4x_4}\left(
\begin{array}{c}
(p_4-ip_3)\Phi_{l}^{\uparrow\uparrow}(r)e^{il\vartheta}\\
-2l\Phi_{l+1}^{\uparrow\downarrow}(r)e^{i(l+1)\vartheta}\\
\lambda \Phi_l^{\uparrow\uparrow}(r)e^{il\vartheta}\\
0
\end{array}
\right)
\end{equation*}
\begin{equation*}
\psi^{(4)}_l=e^{-ip_3x_3-ip_4x_4}\left(
\begin{array}{c}
\frac{\mu^2}{2l}\Phi_l^{\uparrow\uparrow}(r)e^{il\vartheta}\\
(p_4+ip_3)\Phi_{l+1}^{\uparrow\downarrow}(r)e^{i(l+1)\vartheta}\\
0\\
\lambda \Phi_{l+1}^{\uparrow\downarrow}(r)e^{i(l+1)\vartheta}\\
\end{array}
\right).
\end{equation*}

All four spinors are eigenfunctions,
\begin{eqnarray*}
J_3\psi_l^{(m)}=\left(l+\frac{1}{2}\right)\psi_l^{(m)},
\end{eqnarray*}
 of the total momentum projection operator onto $x_3$.
\begin{gather*}
J_i=L_i+S_i,\quad L_i=-i\varepsilon_{ijk}x_j\partial_k,\quad S_i=\frac12\Sigma_i,\\
J_3=-i\frac{\partial}{\partial \vartheta}+\frac{1}{2}\left(\begin{array}{cc}
\sigma_3&0\\
0&\sigma_3
\end{array}\right).
\end{gather*}

Only two of these solutions at given $l$ are linearly independent. We select
\begin{equation*}
\psi_l=A\psi^{(1)}_l+B\psi^{(4)}_l
\end{equation*}
as a  general solution to equation \eqref{dirac} for the reason that $\psi^{(1)}_l$ and $\psi^{(4)}_l$ remain linearly independent in the limit $\lambda\to 0$. The limit will be used in the next section for the solving the  Dirac equation in Minkowski space-time.

Boundary condition \eqref{bagbc} with $\theta=\pi/2$ leads to the equation defining the values of the parameter $\mu$ as well as  the ratio of
 $A$ and $B$. For $l\geqslant 0$ one gets
\begin{widetext}
\begin{eqnarray}
&& A\left(\frac{\mu^2}{2(1+l)}\Phi^{\uparrow\downarrow}_{l+1}(R)+\lambda \Phi^{\uparrow\uparrow}_{l}(R)\right)+B\left(\lambda \Phi^{\uparrow\downarrow}_{l+1}(R)+2(l+1)\Phi^{\uparrow\uparrow}_{l}(R)\right)=0,
\nonumber\\
&&A\Phi^{\uparrow\uparrow}_l(R)+B \Phi^{\uparrow\downarrow}_{l+1}(R)=0.
\nonumber
\end{eqnarray}
 This system has a nontrivial solution for $A$ and $B$ if the determinant of the matrix composed of the coefficients in front of them is equal to zero
\begin{gather}
\label{muquark}
\left[\Phi^{\uparrow\uparrow}_l(R)\right]^2=\left[\frac{\mu}{2(1+l)}\Phi^{\uparrow\downarrow}_
{l+1}(R)\right]^2.
\end{gather}
This equation defines the spectrum of $\mu^2$.  States with definite spin orientation with respect to the chromomagnetic field  are mixed in the boundary condition, and the spin projection onto the direction of the field is not a good quantum number unlike the projection of the total momentum $j_3$ as it is taken into account in Fig.\ref{Fig:Q_eigenv}. As is illustrated in Fig.\ref{Fig:Q_eigenv} there is a discrete set of solutions $\mu_{ilk}>0$  which depend also on the color orientation $\hat u_i$ ($j_3=(2l+1)/2$ with $l\in Z$, $k\in N$, $j=1,2,3$).  
As a rule one can omit the color index $j$ assuming that $\mu_{lk}$ is a diagonal color matrix for any $l,k$. 
The values $\mu^2_{lk}$ has to be used to find the relation between $A$ and $B$
\begin{gather}
\frac{B_{lk}}{A_{lk}}=-\left.\frac{\Phi^{\uparrow\uparrow}_l(R)}{\Phi^{\uparrow\downarrow}_{l+1}(R)}\right|_{\mu^2=\mu^2_{lk}}= (-1)^{k+1}\frac{\mu_{lk}}{2(l+1)},
\label{lambdapm}\\
\lambda_{lk}=\pm\sqrt{\mu_{lk}^2+p_3^2+p_4^2}=\pm |\lambda_{lk}|.
\nonumber
\end{gather}
Here $\mu_{lk}$ is taken to be positive, and $\lambda_{lk}$ takes both positive and negative values.  Equation \eqref{muquark} has been used in combination with observation (by inspection) that the sign of the ratio $B_{lk}$ and  $A_{lk}$  depends on $k\in N$ as it is indicated in \eqref{lambdapm} irrespectively to $l$ and color orientation.

Analogous consideration for the case $l<0$ leads to the equation for $\mu$
\begin{gather*}
\left[\Phi^{\uparrow\downarrow}_{l+1}(R)\right]^2=\left[\frac{\mu}{2l}\Phi^{\uparrow\uparrow}_
{l}(R)\right]^2,
\end{gather*}
and for the ratio of coefficients
\begin{gather*}
\frac{A_{lk}}{B_{lk}}=-\left.\frac{\Phi^{\uparrow\uparrow}_l(R)}{\Phi^{\uparrow\downarrow}_{l+1}(R)}\right|_{\mu^2=\mu^2_{lk}}= (-1)^{k}\frac{\mu_{lk}}{2l},
\label{lumpm}\\
\lambda_{lk}=\pm\sqrt{\mu_{lk}^2+p_3^2+p_4^2}=\pm |\lambda_{lk}|.
\nonumber
\end{gather*}

The orthogonal normalized set of solutions has the form for $l\geqslant 0$
\begin{equation*}
\psi_{lk}^{(\pm)}=\frac{A_{lk}}{(2\pi)^\frac32\sqrt{2|\lambda_{lk}|}}\left(
\begin{array}{c}
\frac{\pm|\lambda_{lk}|+(-1)^{k+1}\mu_{lk}}{\sqrt{p_4+ip_3}} 
\Phi_l^{\uparrow\uparrow}(r)e^{il\vartheta}\\
(-1)^{k+1}\frac{\mu_{lk}\sqrt{p_4+ip_3}}{2(l+1)}\Phi_{l+1}^{\uparrow\downarrow}(r)e^{i(l+1)\vartheta}\\
\sqrt{p_4+ip_3}\Phi_l^{\uparrow\uparrow}(r)e^{il\vartheta}\\
\frac{\mu_{lk}(\mu_{lk}\pm(-1)^{k+1}|\lambda_{lk}|)}{2(l+1)\sqrt{p_4+ip_3}}\Phi_{l+1}^{\uparrow\downarrow}(r)e^{i(l+1)\vartheta}
\end{array}
\right)e^{-ip_3x_3-ip_4x_4},
\label{psi_eucl1}
\end{equation*}
and for $l<0$
\begin{equation*}
\psi_{lk}^{(\pm)}=\frac{B_{lk}}{(2\pi)^\frac32\sqrt{2|\lambda_{lk}|}}\left(
\begin{array}{c}
  \frac{\mu_{lk}(\mu_{lk}\pm(-1)^k|\lambda_{lk}|)}{2l\sqrt{p_4+ip_3}} 
\Phi_l^{\uparrow\uparrow}(r)e^{il\vartheta}\\
\sqrt{p_4+ip_3}\Phi_{l+1}^{\uparrow\downarrow}(r)e^{i(l+1)\vartheta}\\
(-1)^k\frac{\mu_{lk}\sqrt{p_4+ip_3}}{2l}\Phi_l^{\uparrow\uparrow}(r)e^{il\vartheta}\\
  \frac{\pm|\lambda_{lk}|+(-1)^k\mu_{lk}}{\sqrt{p_4+ip_3}}\Phi_{l+1}^{\uparrow\downarrow}(r)e^{i(l+1)\vartheta}
\end{array}
\right)e^{-ip_3x_3-ip_4x_4}.
\nonumber
\end{equation*}
The spinors  $\psi_{lk}^{(+)}$ and $\psi_{lk}^{(-)}$ correspond to the 
positive and negative eigenvalues $\lambda_{lk}$ in \eqref{lambdapm}  
respectively, they are eigenfunctions of $J_3$ with $j_3=l+1/2$. 
Normalization constants are
\begin{eqnarray*}
A^{-2}_{jlk}(R)&=&\int_0^R 
drr\left[\left(\frac{\mu_{jlk}}{2(l+1)}\Phi_{l+1}^{\uparrow\downarrow}(r)\right)^2+\left(\Phi_{l+1}^{\uparrow\uparrow}(r)\right)^2\right]
\nonumber\\
B^{-2}_{jlk}(R)&=&\int_0^R 
drr\left[\left(\frac{\mu_{jlk}}{2l}\Phi_{l+1}^{\uparrow\uparrow}(r)\right)^2+\left(\Phi_{l+1}^{\uparrow\downarrow}(r)\right)^2\right]
\label{normconst}
\end{eqnarray*}

The same procedure applied to the equation
\begin{equation*}
\bar\psi(x)\stackrel{\leftarrow}{\!\not\!\!D}=\lambda\bar\psi(x)
\end{equation*}
leads to the solutions for $l\geqslant 0$
\begin{equation}
\bar\psi_{lk}^{(\pm)}=\frac{A_{lk}}{(2\pi)^\frac32\sqrt{2|\lambda_{lk}|}}\left(
\begin{array}{c}
\pm\sqrt{p_4+ip_3}\Phi^{\uparrow\uparrow}_l(r)e^{-il\vartheta}\\
\frac{\mu_{lk}(\mp\mu_{lk}+(-1)^{k+1}|\lambda_{lk}|)}{2(1+l)\sqrt{p_4+ip_3}} 
\Phi^{\uparrow\downarrow}_{l+1}(r)e^{-i(l+1)\vartheta}\\
\frac{\pm(-1)^{k}\mu_{lk}+|\lambda_{lk}|}{\sqrt{p_4+ip_3}}\Phi^{\uparrow\uparrow}_l(r)e^{-il\vartheta}\\
\mp(-1)^{k}\frac{\mu_{lk}\sqrt{p_4+ip_3}}{2(1+l)}\Phi^{\uparrow\downarrow}_{l+1}(r)e^{-i(l+1)\vartheta}
\end{array}
\right)^\textrm{T}e^{ip_3x_3+ip_4x_4},
\label{psi_eucl}
\end{equation}
and for $l<0$
\begin{equation}
\bar\psi_{lk}^{(\pm)}=\frac{B_{lk}}{(2\pi)^\frac32\sqrt{2|\lambda_{lk}|}}\left(
\begin{array}{c}
\pm(-1)^{k}\frac{\mu_{lk}\sqrt{p_4+ip_3}}{2l}\Phi^{\uparrow\uparrow}_l(r)e^{-il\vartheta}\\
\frac{|\lambda_{lk}|\mp(-1)^k\mu_{lk}}{\sqrt{p_4+ip_3}} 
\Phi^{\uparrow\downarrow}_{l+1}(r)e^{-i(l+1)\vartheta}\\
\frac{\mu_{lk}(\mp\mu_{lk}+(-1)^k|\lambda_{lk}|)}{2l\sqrt{p_4+ip_3}}\Phi^{\uparrow\uparrow}_l(r)e^{-il\vartheta}\\
\pm\sqrt{p_4+ip_3}\Phi^{\uparrow\downarrow}_{l+1}(r)e^{-i(l+1)\vartheta}\\
\end{array}
\right)^\textrm{T}e^{ip_3x_3+ip_4x_4}
.
\nonumber
\end{equation}

\begin{figure}[h!]
 \centerline{\includegraphics[width=80mm,angle=0]{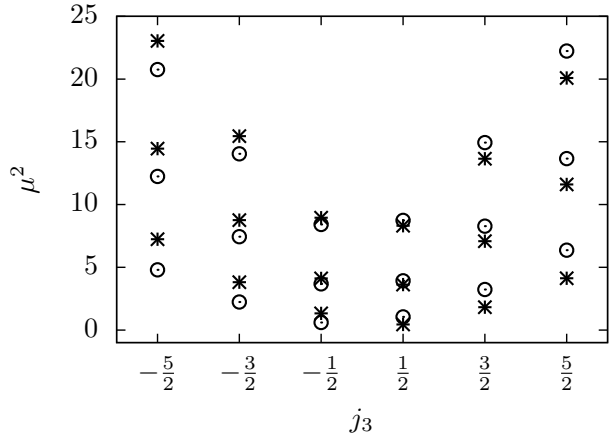}}
    \caption{ The lowest values of $\mu$ solving Eq.~\eqref{muquark} for  $\sqrt{H}R=1.6$. Here $j_3=l+1/2$ is the projection of the total momentum on the direction of the chromomagnetic field. 
Eigenvalues are denoted by asterisks in the case of positive $u_j$ and 
by circles in the case of negative $u_j$.}
   
 \label{Fig:Q_eigenv}
\end{figure}

\subsection{ Quasiparticles} 

To get insight into the physical treatment of above-considered Euclidean eigenmodes  one has to solve  the Minkowski space Klein-Gordon and Dirac equations in the presence of chromomagnetic field inside the cylinder with the bag-like boundary conditions. Solutions
describe the  elementary quasiparticle excitations inside the thick cylindrical domain wall junction. Quite detailed analysis of the notion of quasiparticles in relativistic quantum field theory can be found in  \cite{Arteaga:2008ux}. Unlike the fundamental elementary and composite particles the characteristic properties  of quasiparticles (for instance the specific form of the dispersion relation) need not be necessarily Lorentz invariant or even gauge invariant. The overall statement of the problem under consideration  necessarily assumes that space direction along the chromomagnetic field is singled out by underlining experimental setup as it coincides with the direction of the strong magnetic field generated for short time in heavy ion collision.  In generic relativistic frame both chromoelectric and chromomagnetic fields are present inside the domain wall junction. However since the topological charge density vanishes in the region (see Fig.\ref{Fig:chromo_bag}) there exists specific frame where chromoelectric field is absent. This frame is the most convenient for our purposes.

\subsubsection{Adjoint representation: color charged bosons} 

In Minkowski space-time the problem \eqref{cylinder_s} and \eqref{HDC} 
turns to the wave equation 
\begin{eqnarray*}
\label{cylinder_s_m}
-\left(\partial_\mu-i\breve B_\mu\right)^2\phi(x)=0
\end{eqnarray*}
for color charged adjoint spin zero field inside a cylindrical wave guide. 
As it follows from \eqref{n_diag} and futher discussion the charged components of the adjoint field of the color matrix $\breve n$ comes in complex conjugate 
pairs. For instance if $\xi=\pi/6$ then there are two pairs $\phi_1\pm i\phi_2$ and $\phi_4\pm i\phi_5$. Thus $\phi^a$ is a complex scalar field,
the corresponding solution  of \eqref{cylinder_s_m} satisfying boundary condition \eqref{HDC} takes the form
\begin{eqnarray}
&&\phi^a(x)=\sum_{lk}\int\limits_{-\infty}^{+\infty}\frac{dp_3}{2\pi}\frac{1}{\sqrt{2\omega_{alk}}} \left[a^{+}_{akl}(p_3)e^{ix_0\omega_{akl}-ip_3x_3}
+b_{akl}(p_3)e^{-ix_0\omega_{akl}+ip_3x_3}\right]e^{il\vartheta}\phi_{alk}(r),
\label{seigenf_mink}
\\
&&\phi^{a\dagger}(x)=\sum_{lk}\int\limits_{-\infty}^{+\infty}\frac{dp_3}{2\pi}\frac{1}{\sqrt{2\omega_{alk}}} \left[b^{+}_{akl}(p_3)e^{-ix_0\omega_{akl}+ip_3x_3}
+a_{akl}(p_3)e^{ix_0\omega_{akl}-ip_3x_3}\right]e^{-il\vartheta}\phi_{alk}(r),
\nonumber
\\
&&p_0^2=p_3^2+\mu_{akl}^2, 
\nonumber\\ 
&&p_0=\pm\omega_{akl}(p_3), \ \omega_{akl}=\sqrt{p_3^2+\mu_{akl}^2},
\label{mass_mink}
\\
&& k=0,1,\dots,\infty, \ \ l\in Z,
\nonumber
\end{eqnarray}
with $\phi_{alk}(r)$ defined in \eqref{scalarphi} but here it is assumed to be normalized
\begin{eqnarray*}
\int\limits_0^\infty dr r\int\limits_0^{2\pi} d\vartheta  e^{i(l-l')\vartheta} \phi_{alk}(r)  \phi_{al'k'}(r)=\delta_{ll'}\delta_{kk'}. 
\end{eqnarray*}
Equation \eqref{mass_mink} can be treated as the dispersion relation between energy $p_0$ and momentum $p_3$ for the quasiparticles with masses $\mu_{akl}$.   These quasiparticles are extended in $x_1$ and $x_2$ directions and are classified by the quantum numbers $l,k$.   
The orthogonality, normalization and completeness of the set of
functions $e^{il\vartheta}\phi_{alk}(r)$  guarantees the standard canonical commutation relations for the field $\phi^a$ and its canonically conjugated momentum if $a^\dagger_{akl}(p_3)$, $a_{akl}(p_3)$,  $b^\dagger_{akl}(p_3)$ and $b_{akl}(p_3)$ are assumed to satisfy the standard commutation relations for creation and annihilation operators. The Fock space of states for the quasiparticles with masses $\mu_{akl}$ can be constructed by means of the standard QFT methods. 
This treatment provides one with a suitable terminology  and formalism for discussion of the confining properties of various gluon field configurations in the context of QFT:
unlike the chromomagnetic field the (anti-)self-dual fields characteristic for the bulk of domain network configuration (see the LHS plot in Fig.~\ref{Fig:kink_network}) lead to purely discrete spectrum of eigenmodes in Euclidean space and do not possess any quasiparticle treatment in terms of  dispersion relation between energy and momentum for elementary  color charged excitations. If there is a reason for long-lived defect in the form of thick domain wall junction then its boundary defines a shape and a size 
for the space region which can be populated by color charged quasiparticles. 

The vector adjoint field can be elaborated in the similar to the scalar case way. A modification relates just to the inclusion of polarization vectors. 
As it has already been mentioned the most important feature is the absence  of tachyonic mode of the vector color charged field if $R<R_{\mathrm c}$.
Disappearance of the tachyonic mode for subcritical size of the trap is one of the most important observations of this paper.

\subsubsection{Fundamental representation: color charged fermions}
 
Neither the background field nor the boundary condition involve the time coordinate. The solution of the Dirac equation 
\begin{eqnarray*}
&&i\!\not\!\!D\psi(x)=0,
\label{dirac1}
\end{eqnarray*}
satisfying condition
\eqref{bagbc} 
can be obtained from Euclidean solutions \eqref{psi_eucl} (unnormalized solutions have to be used) by the analytical continuation $p_4\to ip_0$, $x_4\to ix_0$ and the requirement $\lambda_{lk}=0$,
which leads to the energy-momentum relation for the solutions with definite $j_3$, $k$ and color $j$
\begin{gather*}
p_0^2=p_3^2+\mu_{jlk}^2 ,\ \ 
p_0=\pm\omega_{jlk}(p_3), 
\\ \omega_{jlk}=\sqrt{p_3^2+\mu_{jlk}^2}.
\end{gather*}
 Finally the solution of the Dirac equation takes the form
\begin{eqnarray*}
\psi^{j}(x)=\sum_{lk}\int\limits_{-\infty}^{+\infty}\frac{dp_3}{2\pi}\frac{1}{\sqrt{2\omega_{jlk}}}
\left[a^{\dagger}_{jlk}(p_3)\chi_{jlk}(p_3|r,\vartheta) 
e^{ix_0\omega_{jlk}-ix_3p_3}+b_{jlk}(p_3)\upsilon_{jlk}(p_3|r,\vartheta)e^{-ix_0\omega_{jlk}+ix_3p_3}\right],
\\
\bar\psi^{j}(x)=\sum_{lk}\int\limits_{-\infty}^{+\infty}\frac{dp_3}{2\pi}\frac{1}{\sqrt{2\omega_{jlk}}}
\left[b^{\dagger}_{jlk}(p_3)\bar\chi_{jlk}(p_3|r,\vartheta) 
e^{-ix_0\omega_{jlk}+ix_3p_3}+a_{jlk}(p_3){\bar\upsilon																}_{jlk}(p_3|r,\vartheta)e^{ix_0\omega_{jlk}-ix_3p_3}\right].
\end{eqnarray*}
Here the pair of spinors for positive $\chi_{lk}$ and negative $\upsilon_{lk}$ energy solutions are 
\begin{eqnarray*}
\chi_{lk}=A_{lk}\left(
\begin{array}{c}
(-1)^{k+1}\frac{\mu_{lk}}{\sqrt{\omega_{lk}+p_3}} \Phi_l^{\uparrow\uparrow}(r)e^{il\vartheta}\\
 i(-1)^{k+1}\frac{\mu_{lk}\sqrt{\omega_{lk}+p_3}}{2(l+1)}\Phi_{l+1}^{\uparrow\downarrow}(r)e^{i(l+1)\vartheta}\\
i\sqrt{\omega_{lk}+p_3}\Phi_l^{\uparrow\uparrow}(r)e^{il\vartheta}\\
\frac{\mu^2_{lk}}{2(l+1)\sqrt{\omega_{lk}+p_3}}\Phi_{l+1}^{\uparrow\downarrow}(r)e^{i(l+1)\vartheta}
\end{array}
\right),
\ \ \ \ 
\upsilon_{lk}=A_{lk}\left(
\begin{array}{c}
(-1)^{k+1}\frac{\mu_{lk}}{\sqrt{\omega_{lk}+p_3}} \Phi_l^{\uparrow\uparrow}(r)e^{il\vartheta}\\
 i(-1)^{k}\frac{\mu_{lk}\sqrt{\omega_{lk}+p_3}}{2(l+1)}\Phi_{l+1}^{\uparrow\downarrow}(r)e^{i(l+1)\vartheta}\\
-i\sqrt{\omega_{lk}+p_3}\Phi_l^{\uparrow\uparrow}(r)e^{il\vartheta}\\
\frac{\mu^2_{lk}}{2(l+1)\sqrt{\omega_{lk}+p_3}}\Phi_{l+1}^{\uparrow\downarrow}(r)e^{i(l+1)\vartheta}
\end{array}
\right),
\label{barchi_mink}
\end{eqnarray*}
for $l\geqslant0$ and
\begin{eqnarray*}
\chi_{lk}=B_{lk}\left(
\begin{array}{c}
\frac{\mu^2_{lk}}{2l\sqrt{\omega_{lk}+p_3}} \Phi_l^{\uparrow\uparrow}(r)e^{il\vartheta}\\
 i\sqrt{\omega_{lk}+p_3}\Phi_{l+1}^{\uparrow\downarrow}(r)e^{i(l+1)\vartheta}\\
i(-1)^k\frac{\mu_{lk}\sqrt{\omega_{lk}+p_3}}{2l}\Phi_l^{\uparrow\uparrow}(r)e^{il\vartheta}\\
(-1)^k\frac{\mu_{lk}}{\sqrt{\omega_{lk}+p_3}}\Phi_{l+1}^{\uparrow\downarrow}(r)e^{i(l+1)\vartheta}
\end{array}
\right),
\ \ \ \ 
\upsilon_{lk}=B_{lk}\left(
\begin{array}{c}
\frac{\mu^2_{lk}}{2l\sqrt{\omega_{lk}+p_3}} \Phi_l^{\uparrow\uparrow}(r)e^{il\vartheta}\\
 -i\sqrt{\omega_{lk}+p_3}\Phi_{l+1}^{\uparrow\downarrow}(r)e^{i(l+1)\vartheta}\\
i(-1)^{k+1}\frac{\mu_{lk}\sqrt{\omega_{lk}+p_3}}{2l}\Phi_l^{\uparrow\uparrow}(r)e^{il\vartheta}\\
(-1)^k\frac{\mu_{lk}}{\sqrt{\omega_{lk}+p_3}}\Phi_{l+1}^{\uparrow\downarrow}(r)e^{i(l+1)\vartheta}
\end{array}
\right)
\label{barchi_mink_ln}
\end{eqnarray*}
for $l<0$.  The spinors are normalized as
\begin{eqnarray*}
\int\limits_{0}^{2\pi}d\vartheta\int\limits_0^R dr r 
\chi^\dagger_{jlk}(p_3|r,\vartheta)\chi_{jlk}(p_3|r,\vartheta)=
\int\limits_{0}^{2\pi}d\vartheta\int\limits_0^R dr r 
\upsilon^\dagger_{jlk}(p_3|r,\vartheta)\upsilon_{jlk}(p_3|r,\vartheta)= 2\omega_{jlk}
\end{eqnarray*}

The Dirac conjugated spinors are
\begin{equation*}
\bar\psi^{j}(x)=\psi^{j\dagger}(x)\gamma_0
\end{equation*}
as usual. The Fock space can be constructed by means of the creation and annihilation operators $\left\{ a^{\dagger}_{jlk}(p_3), a_{jlk}(p_3), b^{\dagger}_{jlk}(p_3), b_{jlk}(p_3)\right\}$ satisfying the standard 
anticommutation relations. The one-particle state is characterized by a 
color orientation $j$, momentum $p_3$, projection $j_3=(l+1/2)$ of the total angular momentum and the energy $\omega_{jlk}=\sqrt{p_3^2+\mu^2_{jlk}}$. Since the boundary condition mixes
the states with spin parallel and anti-parallel to the chromomagnetic field the spin projection is not a good quantum number unlike the half-integer valued projection of the total angular momentum $j_3$.

\end{widetext}

\section{Discussion}

 An ensemble of confining gluon configurations has been constructed explicitly as a domain wall networks representing the almost everywhere homogeneous Abelian (anti-)self-dual gluon fields. Confinement is understood here as the absence of the color charged wave-like elementary excitations. The dynamical quark confinement occurs in the (four-dimensional) bulk of the domain wall network. Inside the (three-dimensional) domain walls topological charge density vanishes and the color charged quasiparticles can be excited.
 
Under extreme conditions, in particular under the influence of the strong electromagnetic field specific for relativistic heavy ion collisions, a  relatively stable defect in the confining ensemble, a thick domain wall junction, can be formed. Though the scalar gluon condensate is nonzero everywhere  $\langle g^2F^2\rangle\not=0$, the region of defect is characterized by the vanishing topological charge density $\langle |g^2\tilde FF|\rangle$=0 unlike the rest of the space, which indicates the lack of confinement in the junction. The  quark field excitations  inside the junction are represented by the color charged quasiparticles. The spectrum of gluon excitations besides the trapped color charged modes contains also the color neutral with respect to the background field modes.

Almost obvious but important observation is that there exists a critical size $L_{\rm c}$ of the junction beyond which the tachyonic gluon modes emerge in the excitation spectrum and destabilize the defect. The critical size can be related to the value of the gluon condensate $\langle g^2F^2\rangle$
and in the case of the considered in the paper cylindrical trap $L_{\rm c}\approx 1$fm for the standard value of the condensate, see \eqref{rc}. The specific value of the critical size depends on the geometry of the trap but its very existence and its commensurability with a distance of order of $1$fm is a generic feature. This observation underlines the generic necessity of 
accounting for the essentially finite size of the space-time region in which deconfinement may occur.
The reason is that thermodynamic limit does not exist as the system under consideration disappears as soon as the typical size of the space volume exceeds  the critical value.
 Excess of the internal pressure of the trap filled by many charged quasiparticles leads to its expansion and breakdown of stability followed by its disintegration to many smaller traps (or bags), which is reminiscent of the heterophase fluctuations studied in \cite{Yukalov:2013yj}  as well as the dynamics and statistical mechnaics of bags with a surface tension \cite{Bugaev}. 

The dynamics of the color charged quasiparticles as it is described above is strictly one-dimentional in space. This  feature can be a source of the azimuthal asymmetries in heavy ion collisions, similarly to the approach of paper  \cite{Tuchin:2013ie} upto a substitution of the magnetic field by the Abelian chromomagnetic field~(see also \cite{Bali:2013owa}). However it should be noted that the one-dimensional dynamics is a property of the zero-th order approximation based on the quadratic part of the action.  Taking into account interactions between the quasiparticles according to the interaction terms in  the action should certainly  dither the direction of the quasiparticle momenta, leaving just some degree of azimuthal asymmetry.

\section*{ACKNOWLEDGMENTS} We acknowledge fruitful discussions with
V.Toneev, S. Molodtsov, J. Pawlowski, M.Ilgenfritz, A.Dorokhov, 
K.Bugaev S.Vinitsky, G.Efimov, V.Yukalov, A.Efremov, A.Titov.

\addtocontents{toc}{\protect\contentsline{section}{Bibliography}{\thepage}}

\end{document}